\documentclass{aa}
\usepackage{epsfig}
\usepackage{lscape}
\usepackage{aalongtable}
\usepackage{natbib}
\usepackage{graphics}
\usepackage{txfonts}
\usepackage{color}
\usepackage{times}
\begin{document}

\title{SBS 0335-052E+W: deep VLT/FORS+UVES spectroscopy of the pair of
the lowest-metallicity blue compact dwarf galaxies\thanks{Based on 
observations
collected at the European Southern Observatory, Chile, ESO program 
69.C-0203(A), 71.B-0055(A)), 70.B-0717(A) and 68.B-0310(A).}
\thanks{Tables 1 - 8 are only available in electronic form in the online
edition.}
}
\author{Y. I.\ Izotov \inst{1,2}
\and N. G.\ Guseva \inst{1,2}
\and K. J.\ Fricke \inst{3,2}
\and P.\ Papaderos \inst{4,5}}
\offprints{Y.I. Izotov, izotov@mao.kiev.ua}
\institute{          Main Astronomical Observatory,
                     Ukrainian National Academy of Sciences,
                     Zabolotnoho 27, Kyiv 03680,  Ukraine
\and                 
                     Max-Planck-Institute for Radioastronomy, Auf dem H\"ugel 69,
                     53121 Bonn, Germany
\and
                     Institute for Astrophysics, University of 
                     G\"ottingen, Friedrich-Hund-Platz 1, 
                     37077 G\"ottingen, Germany
\and                 
                     Instituto de Astrof\'{\i}sica de Andaluc\'{\i}a (CSIC),
                     Camino Bajo de Hu\'etor 50, Granada E-18008, Spain
\and
                     Department of Astronomy and Space Physics,
                     Uppsala University, Box 515, SE-75120 Uppsala, Sweden
}

\date{Received \hskip 2cm; Accepted}

\abstract
{We present deep archival VLT/FORS1+UVES spectroscopic observations 
of the system of two blue compact dwarf (BCD) galaxies SBS 0335--052E and 
SBS 0335--052W.
}
{Our aim is to derive element abundances in different H {{\sc ii}} regions of
this unique system of galaxies and to study spatial abundance variations.
}
{The electron temperature $T_e$(O {{\sc iii}}) in all H {{\sc ii}} regions, 
except for one, is derived from the 
[O {{\sc iii}}] $\lambda$4363/($\lambda$4959+$\lambda$5007) flux 
ratio. We determine ionic abundances of helium, nitrogen, oxygen, neon, 
sulfur, chlorine, argon and iron. The empirical relations for ionization
correction factors are used to derive total abundances of these elements.
}
{The oxygen abundance in the brighter eastern galaxy varies in the range
7.11 to 7.32 in different H {{\sc ii}} regions supporting previous findings 
and suggesting the presence of oxygen abundance variations on spatial scales of
$\sim$ 1 -- 2 kpc. Good seeing during FORS observations 
allowed us to extract spectra of four H {{\sc ii}} regions in SBS 0335--052W.
The oxygen abundance in the brightest region No.1 of SBS 0335--052W is 
7.22 $\pm$ 0.07, consistent with previous determinations.Three other
H {{\sc ii}} regions are much more metal-poor with an unprecedently low
oxygen abundance of 12 + log O/H = 7.01 $\pm$ 0.07 (region No.2), 
6.98 $\pm$ 0.06 (region No.3), and 6.86 $\pm$ 0.14 (region No.4). These are the
lowest oxygen abundances ever derived in emission-line galaxies, supporting
earlier conclusions that SBS 0335--052W is the lowest-metallicity
emission-line galaxy known. Helium abundances derived for the brightest 
H {{\sc ii}} regions of both galaxies are mutually consistent. 
We derive weighted mean He mass fractions of 0.2485$\pm$0.0012 and 
0.2514$\pm$0.0012 for two different sets of He {{\sc i}} emissivities.
The ratios of neon and sulfur to oxygen abundance are similar to 
the respective ratios obtained for other emission-line
galaxies. On the other hand, the chlorine-to-oxygen abundance ratio in
SBS 0335--052E is lower, while the argon-to-oxygen abundance ratio is higher
than those in other low-metallicity galaxies.
The Fe/O abundance ratios in different regions of SBS 0335--052E 
are among the highest for emission-line galaxies implying that iron is almost 
entirely not depleted onto dust grains despite dust being detected
in  this galaxy in earlier ISO and Spitzer observations. 
The N/O abundance ratio
in both galaxies is slightly higher than that derived for other BCDs with
12 + log O/H $<$ 7.6. This implies that the N/O in extremely metal-deficient
galaxies could increase with decreasing metallicity. 
}
{}

\titlerunning{SBS 0335-052E+W: deep VLT/FORS+UVES spectroscopy}
\authorrunning{Y.I.Izotov et al.}
\maketitle


\section{Introduction \label{intro}}

\begin{figure*}[t]
\begin{picture}(17,6)
\put(0.0,0){{\psfig{figure=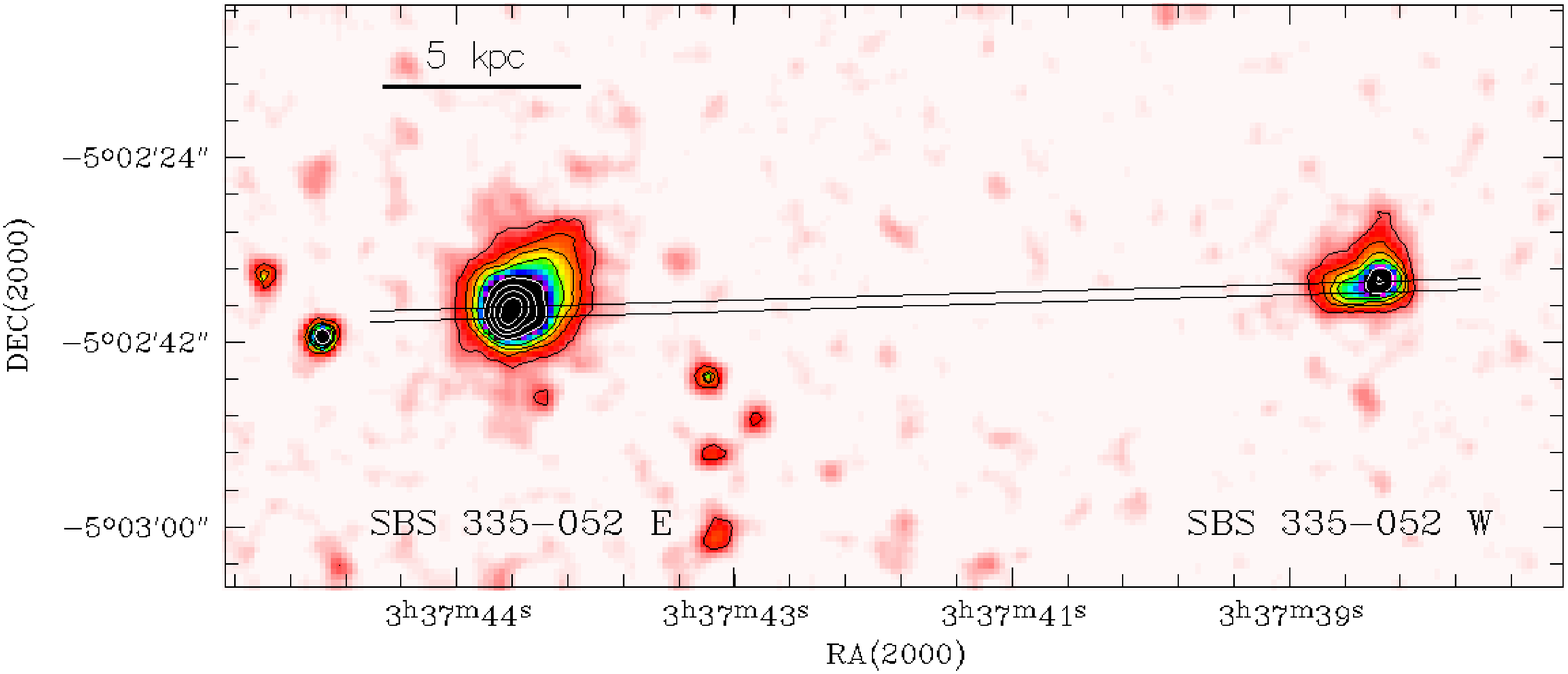,height=6cm,angle=0.,clip=}}}
\put(13.4,0.7){{\psfig{figure=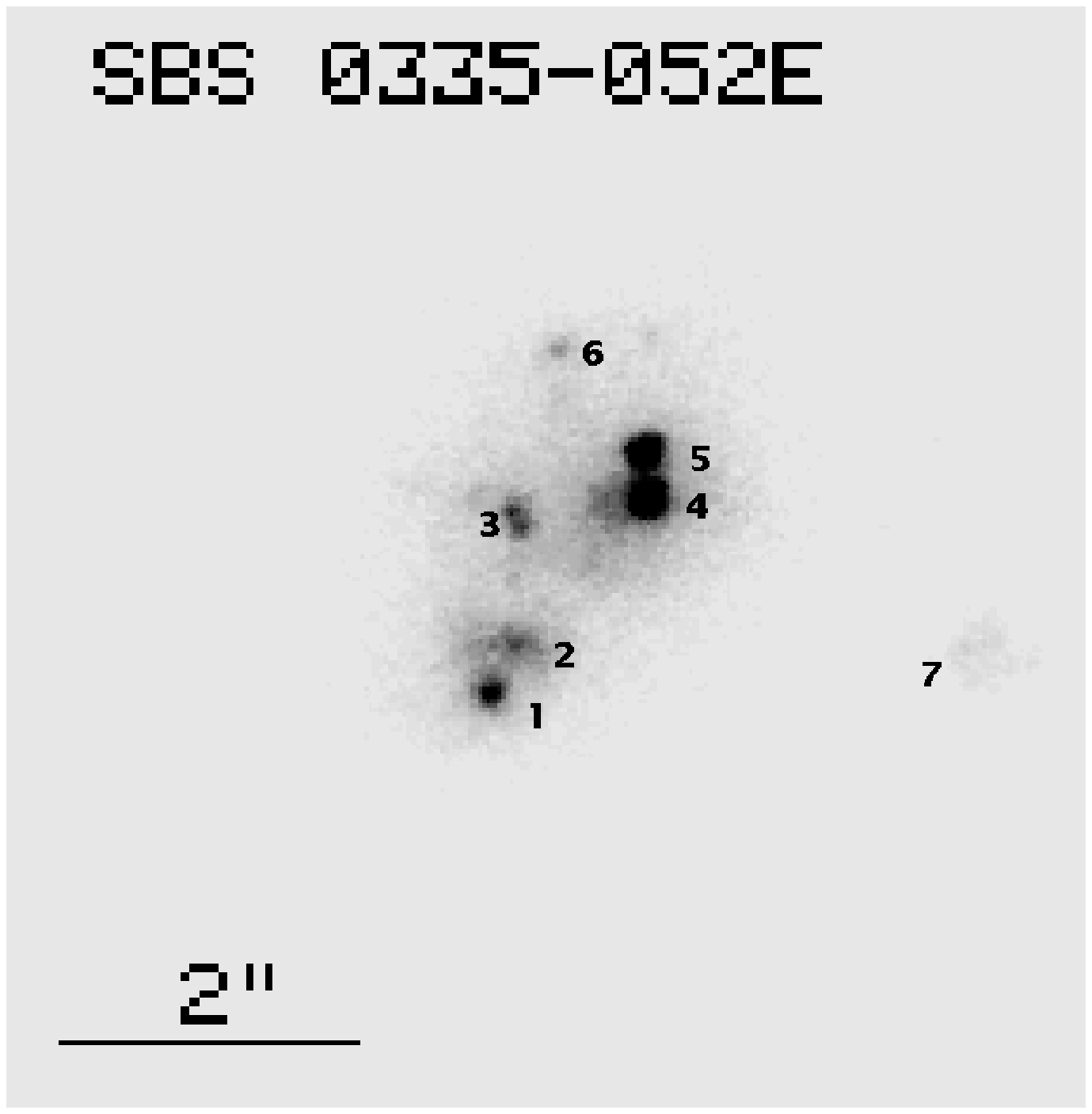,height=5.1cm,angle=0.,clip=}}}
\end{picture}
\caption{{\bf left} 2.2m Calar Alto telescope $B$ image of the
galaxy system SBS 0335--052E and SBS 0335--052W. 
The two straight lines
indicate the location of the slit during VLT/FORS observations.
{\bf right} Archival {\sl HST} UV image of SBS 0335--052E.
The clusters are labelled according to \citet{T97} and \citet{P98}.}
\label{fig1}
\end{figure*}

\begin{figure*}[t]
\hspace*{1.0cm}\psfig{figure=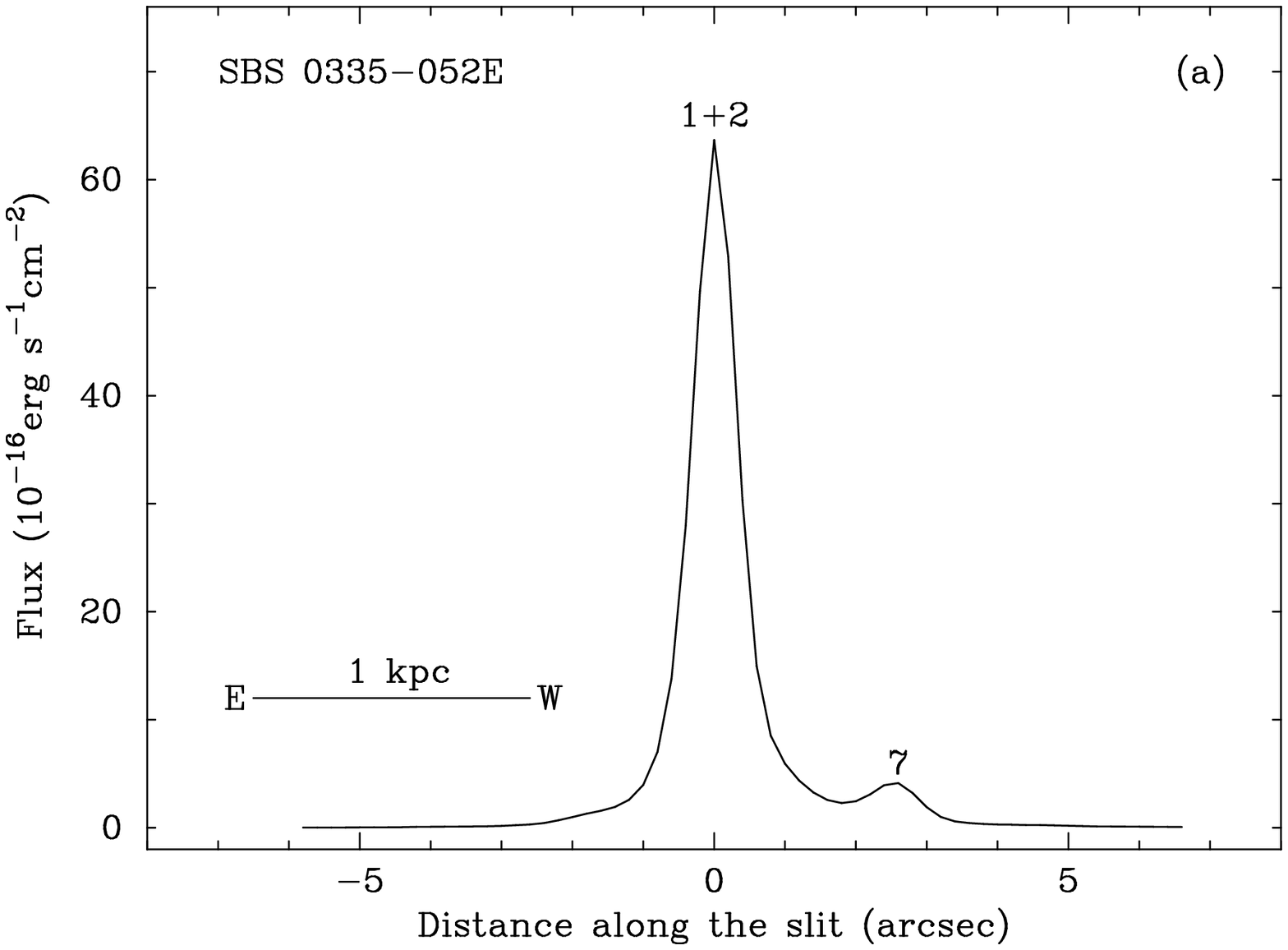,angle=-90,width=7.5cm}
\hspace*{1.0cm}\psfig{figure=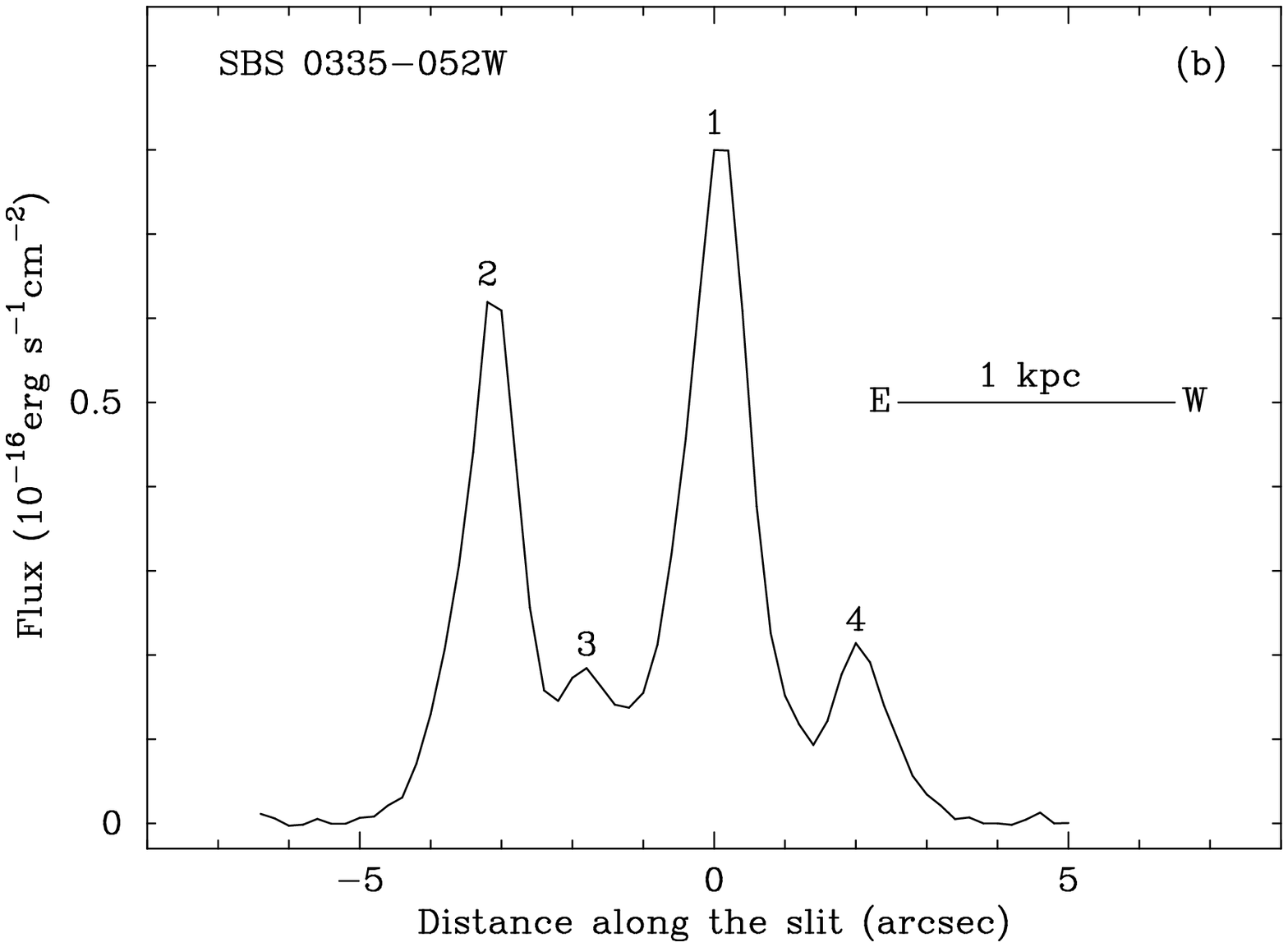,angle=-90,width=7.5cm,clip=}
\caption{H$\beta$ flux distribution along the slit 
(high-resolution FORS observations). 
The orientation of the slit is shown in Fig. \ref{fig1}, left.
The clusters No. 1+2 and No. 7 in SBS 0335--052E are labelled in (a). 
The clusters No. 1 to No. 4  in SBS 0335--052W  are presented in (b).} 
\label{fig2}
\end{figure*}

\begin{figure*}[t]
\hspace*{0.0cm}\psfig{figure=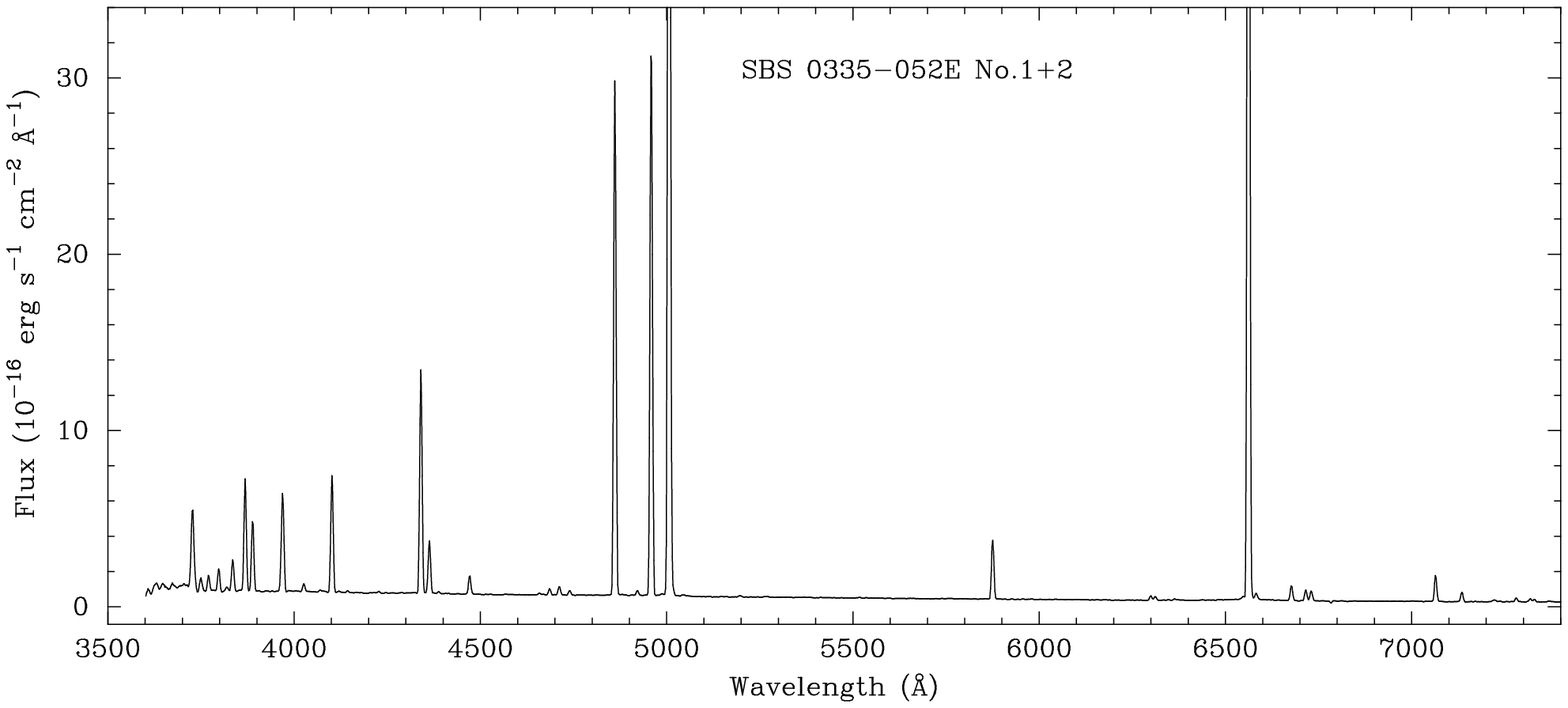,angle=-90,width=8.5cm}
\hspace*{0.4cm}\psfig{figure=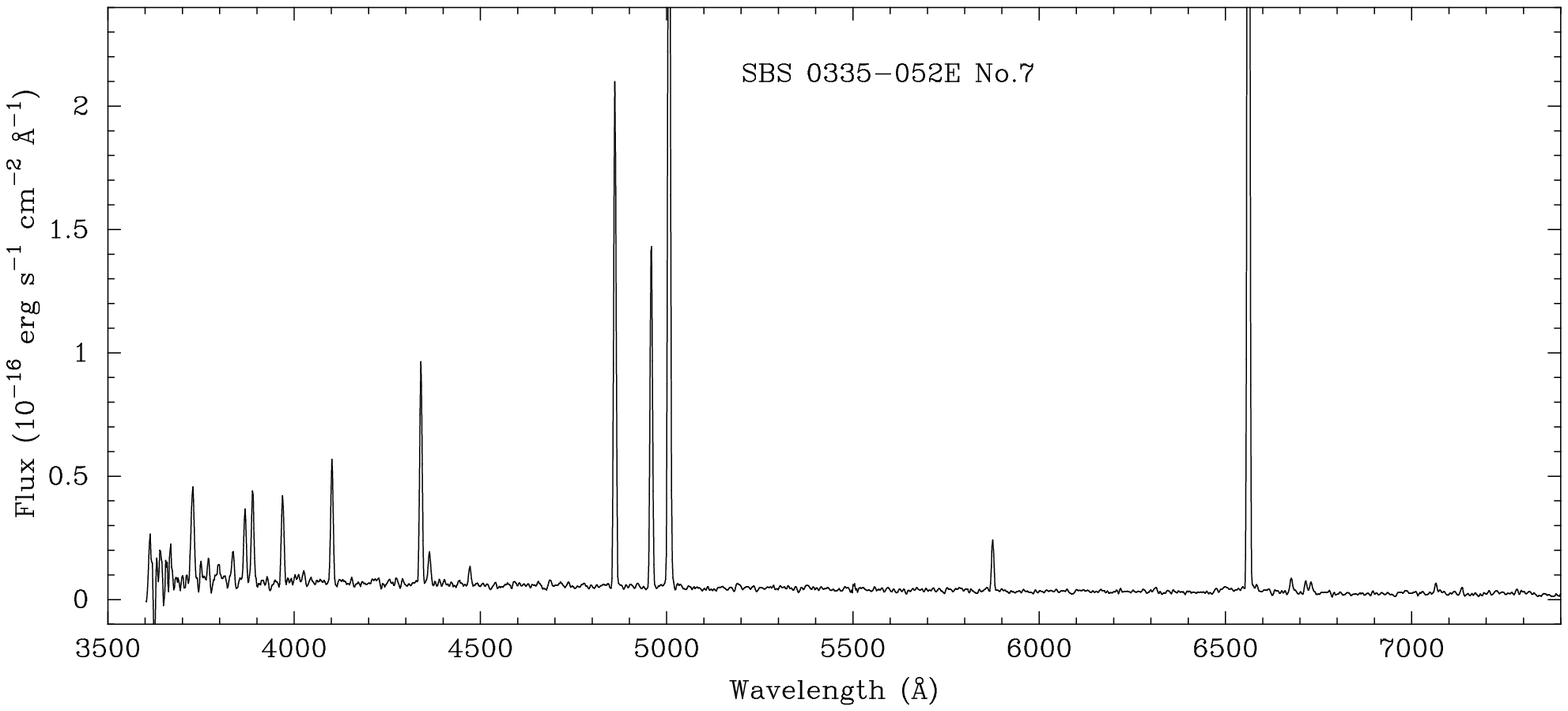,angle=-90,width=8.5cm,clip=}
\hspace*{0.0cm}\psfig{figure=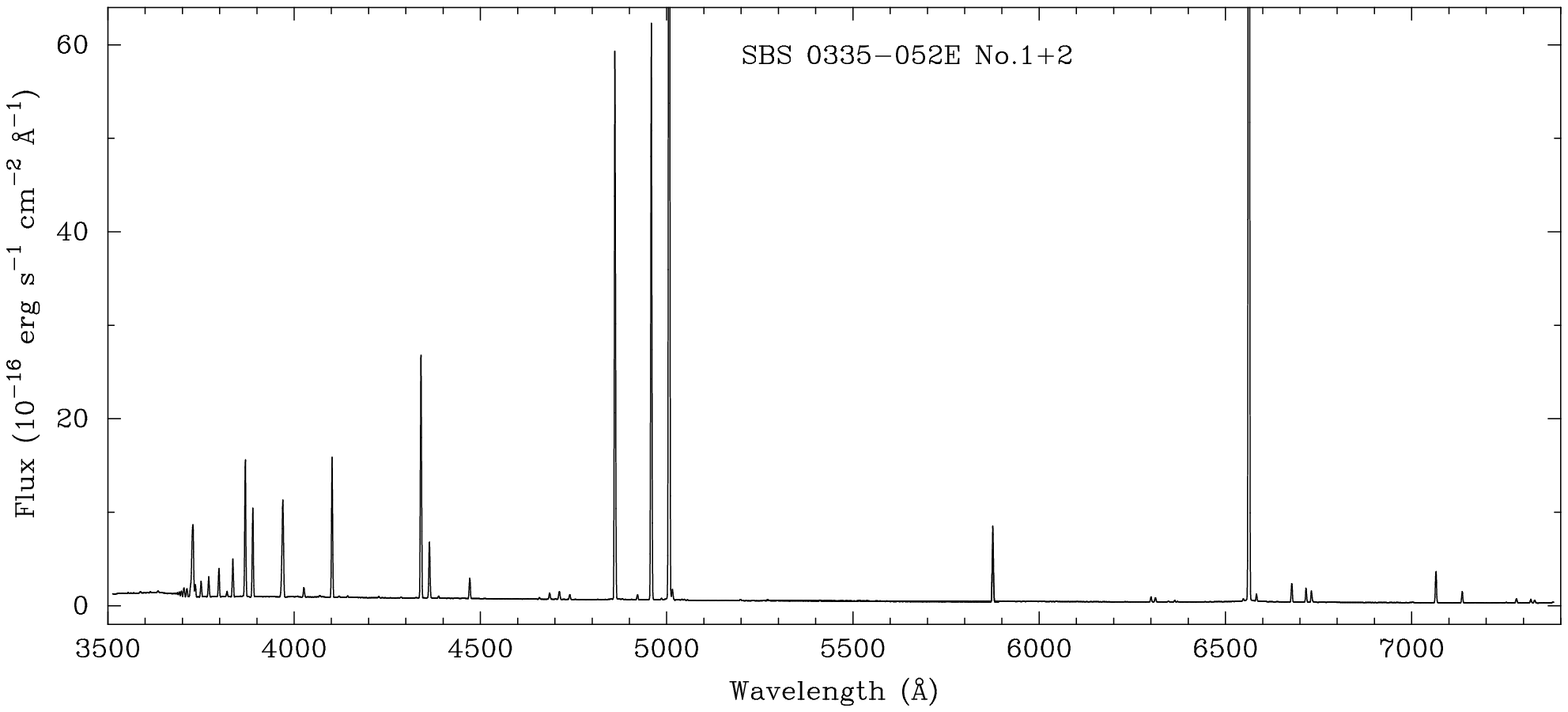,angle=-90,width=8.5cm}
\hspace*{0.4cm}\psfig{figure=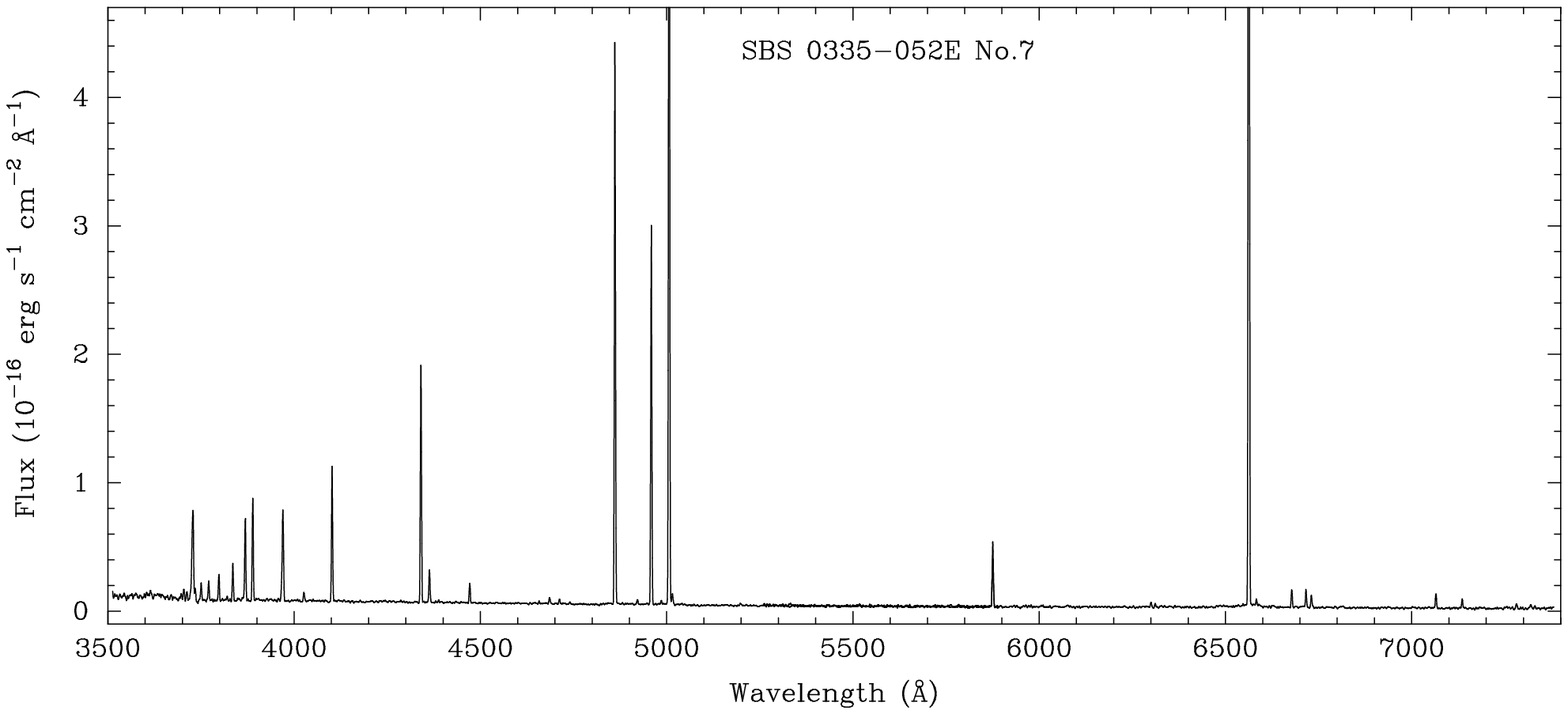,angle=-90,width=8.5cm,clip=}
\caption{Flux-calibrated and redshift-corrected FORS low-resolution 
spectra (upper panel) and FORS high-resolution spectra (lower panel) of 
regions No. 1+2 (left) and No. 7 (right) in SBS 0335--052E 
[ESO program 69.C-0203(A)].}
\label{fig3}
\end{figure*}

\begin{figure*}[t]
\hspace*{0.0cm}\psfig{figure=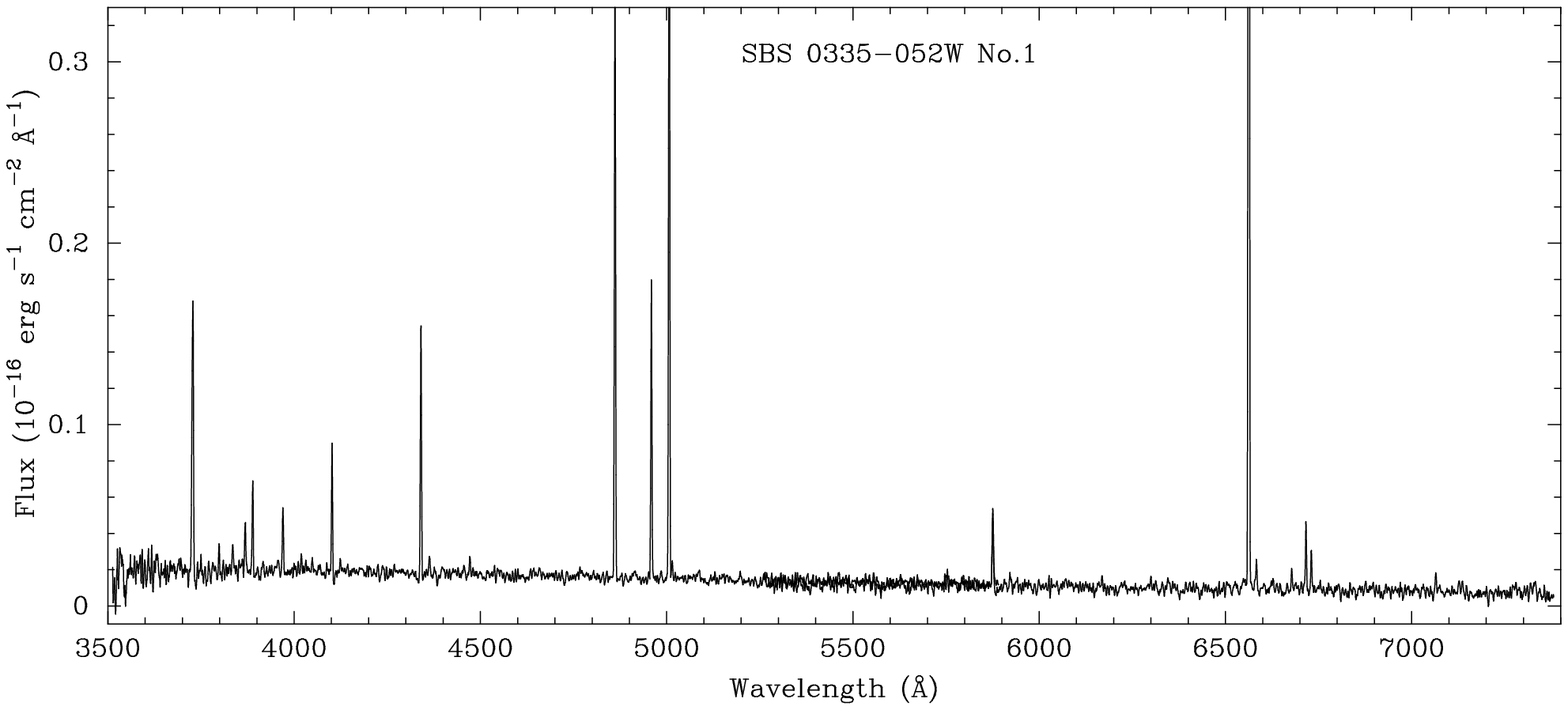,angle=-90,width=8.5cm}
\hspace*{0.4cm}\psfig{figure=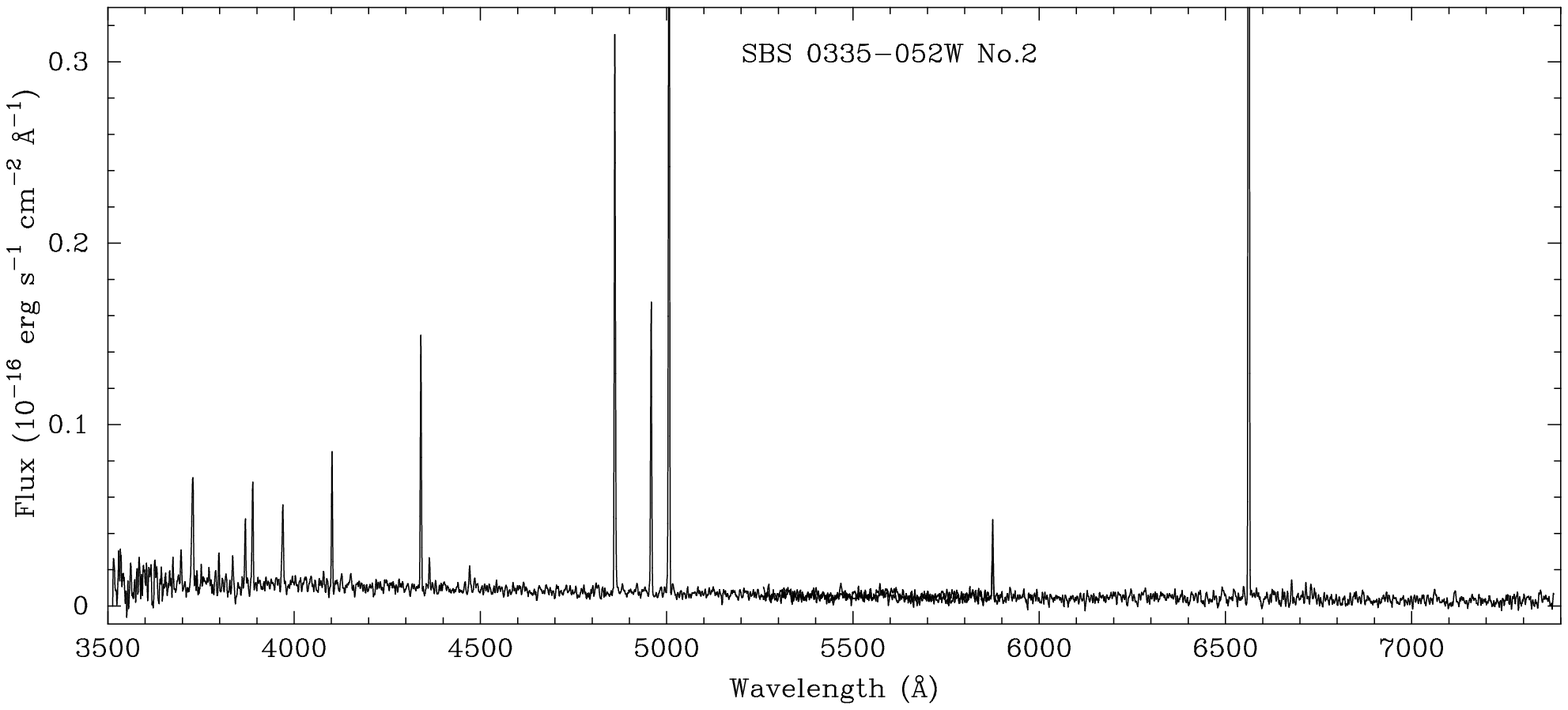,angle=-90,width=8.5cm,clip=}
\hspace*{0.0cm}\psfig{figure=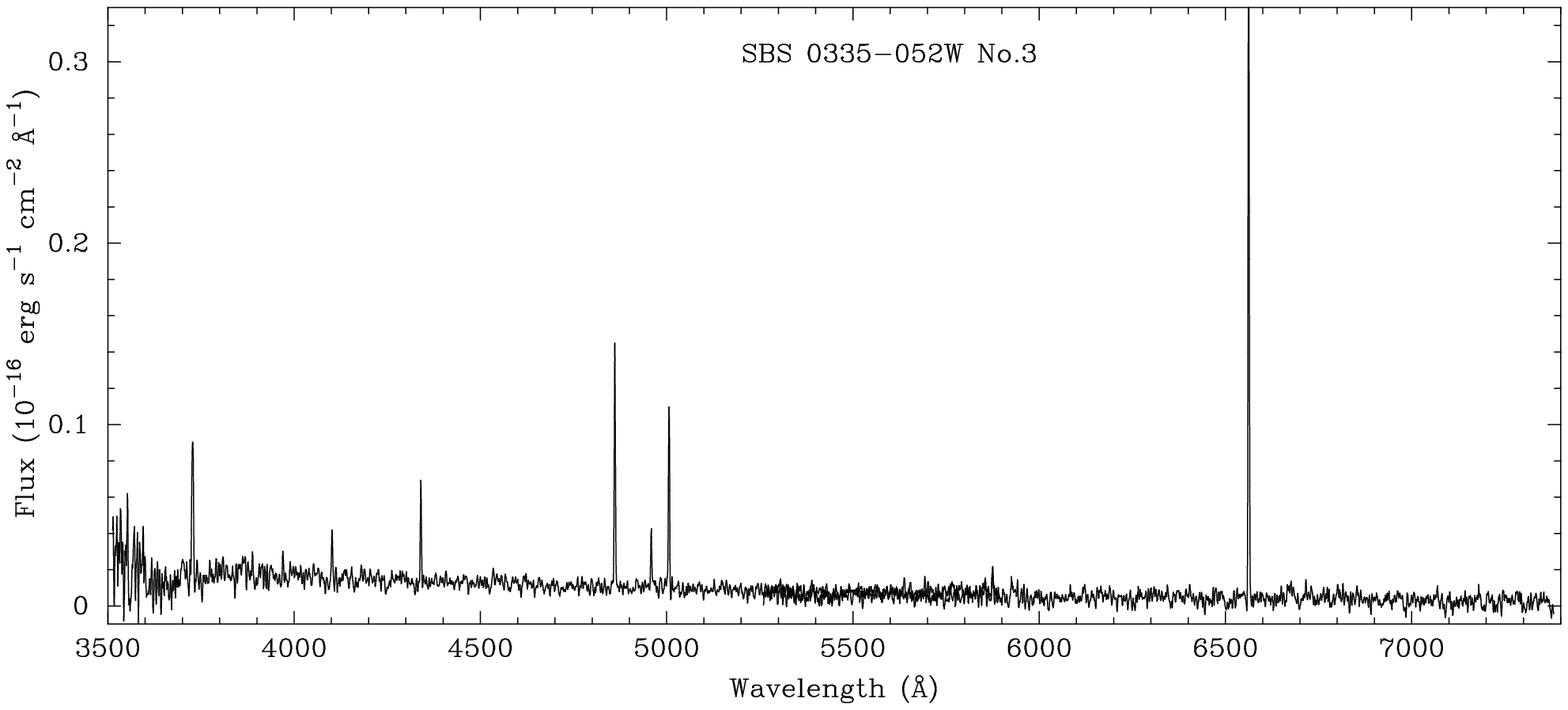,angle=-90,width=8.5cm}
\hspace*{0.4cm}\psfig{figure=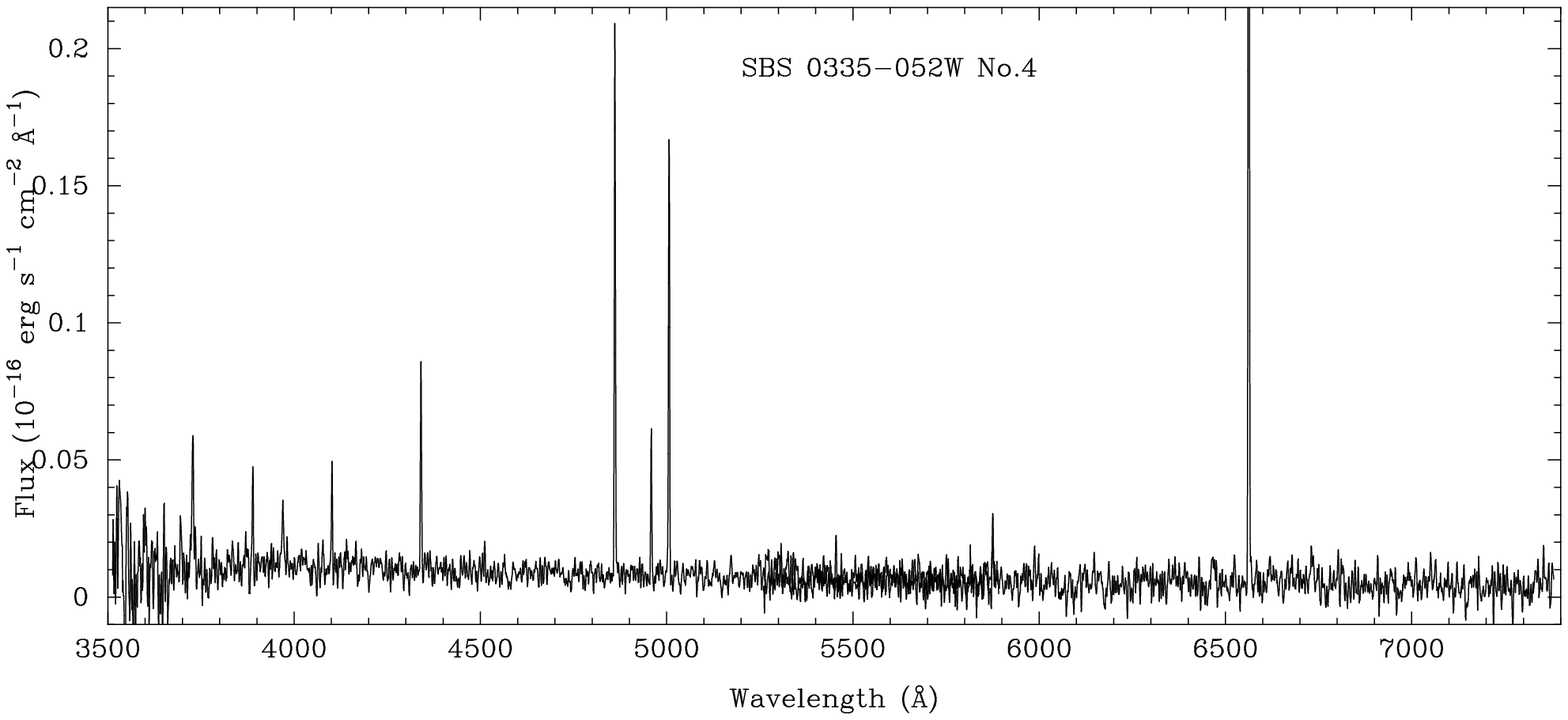,angle=-90,width=8.5cm,clip=}
\caption{Flux-calibrated and redshift-corrected FORS high-resolution 
spectra of regions No. 1 through No. 4 in SBS 0335--052W as labelled
in Fig. \ref{fig2}
[ESO program 69.C-0203(A)].}
\label{fig4}
\end{figure*}

\begin{figure*}[t]
\hspace*{0.0cm}\psfig{figure=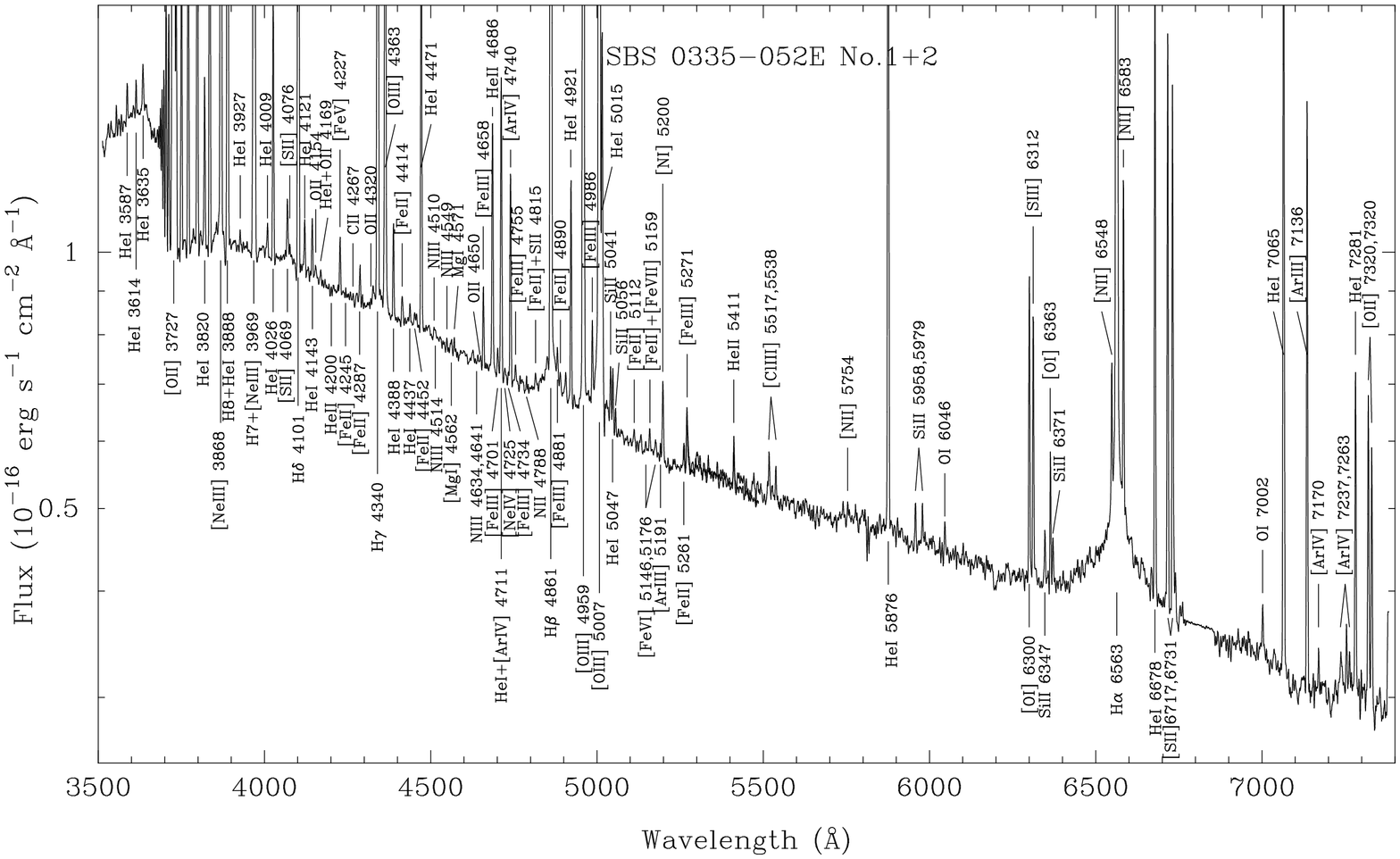,angle=-90,width=18.cm,clip=}
\caption{Flux-calibrated and redshift-corrected FORS high-resolution spectrum 
of regions No. 1+2 (expanded version of Fig. \ref{fig3}, left lower panel,
to visualise weak emission lines more clearly) [ESO program 69.C-0203(A)]. 
Note the presence
of broad emission in the hydrogen lines H$\alpha$, H$\beta$, and H$\gamma$.
No appreciable broad emission is detected in strong forbidden lines, implying
rapid motions of relatively dense ionized gas with an electron number
density $N_e$ $\geq$ 10$^{5-6}$ cm$^{-3}$.}
\label{fig5}
\end{figure*}

\begin{figure*}[t]
\hspace*{0.0cm}\psfig{figure=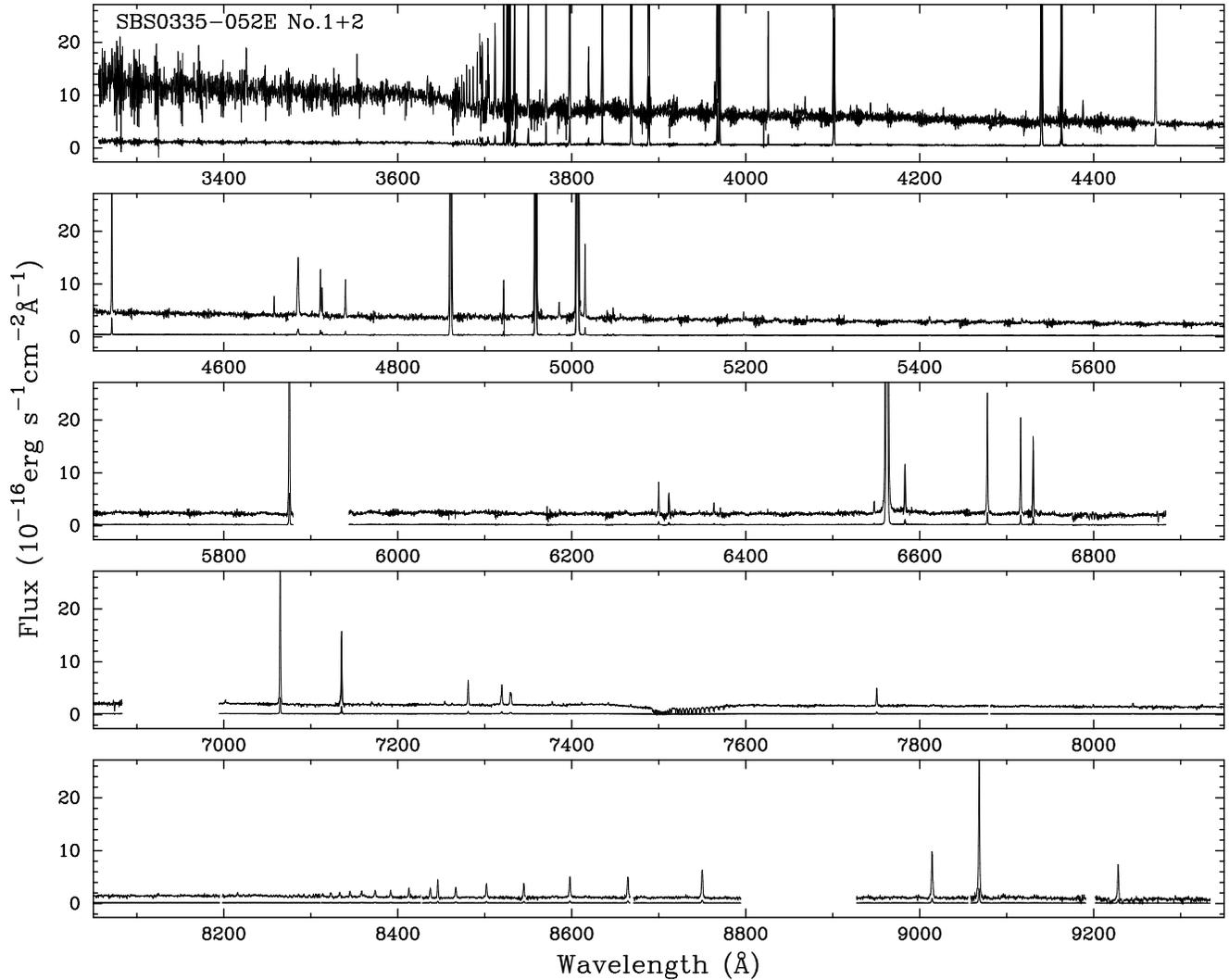,angle=-90,width=17.5cm,clip=}
\caption{Flux-calibrated and redshift-corrected UVES spectrum 
of the regions No. 1+2  in SBS 0335--052E
[ESO program 71.B-0055(A)]. 
}
\label{fig6}
\end{figure*}

\begin{figure*}[t]
\hspace*{0.0cm}\psfig{figure=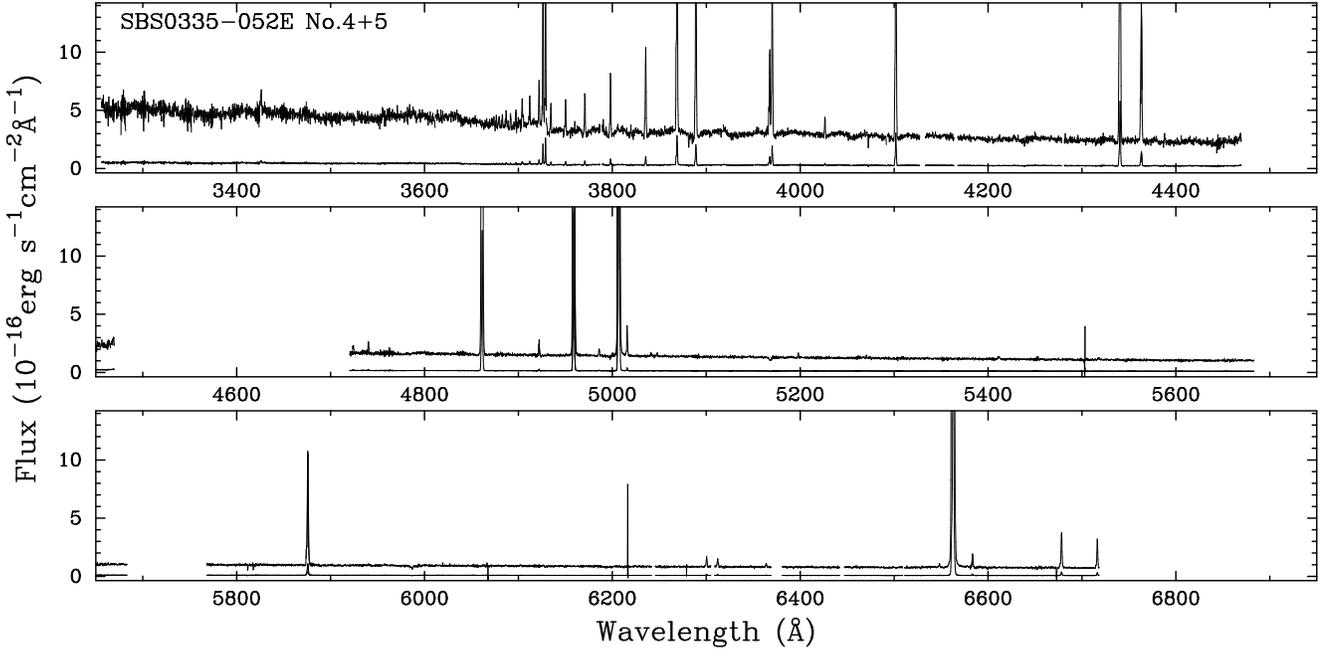,angle=-90,width=17.5cm,clip=}
\caption{Flux-calibrated and redshift-corrected UVES spectrum of 
regions No. 4+5  in SBS 0335--052E [ESO program 70.B-0717(A)]. 
}
\label{fig7}
\end{figure*}

\begin{figure*}[t]
\hspace*{0.0cm}\psfig{figure=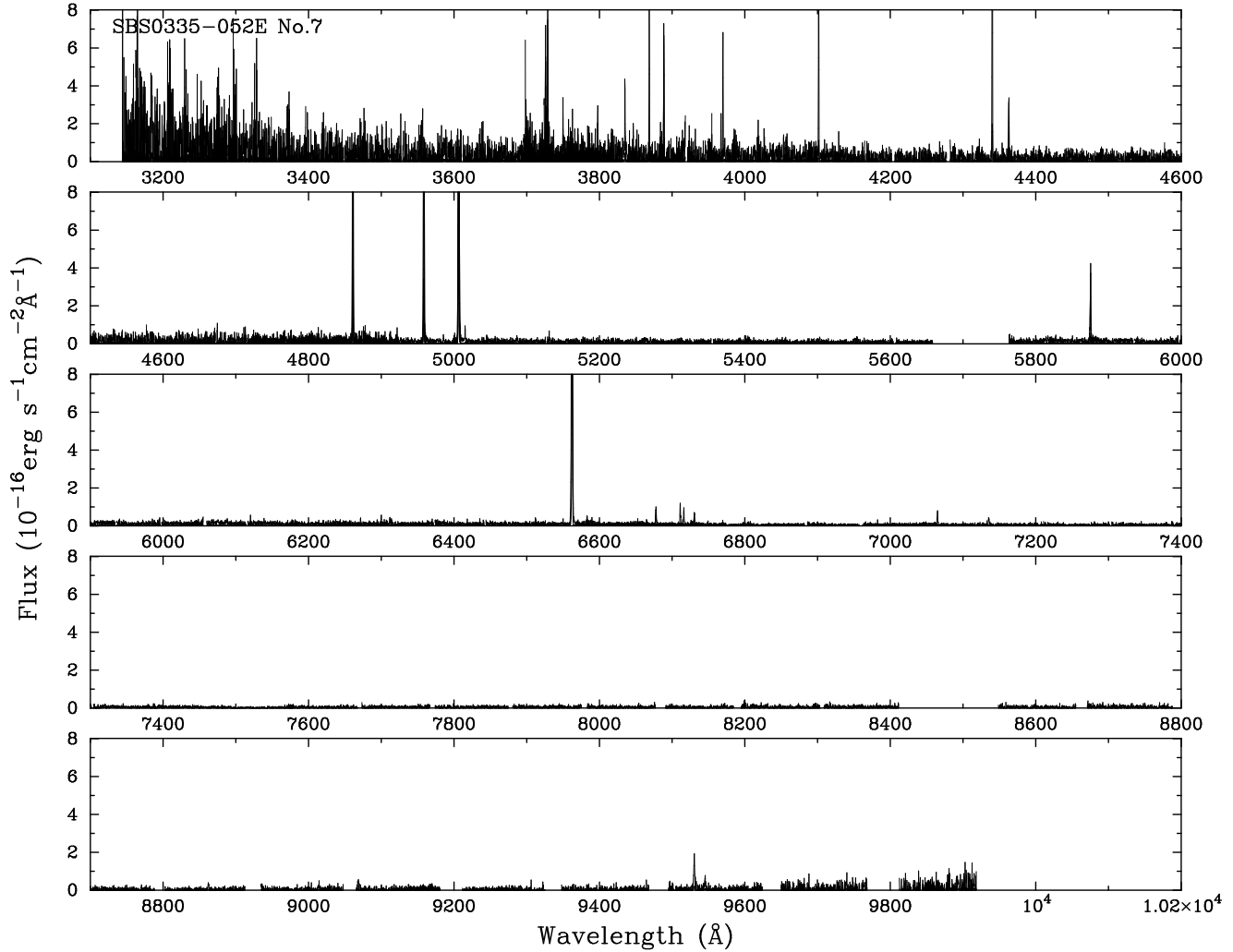,angle=-90,width=17.5cm,clip=}
\caption{Flux-calibrated and redshift-corrected UVES spectrum of 
region No. 7  in SBS 0335--052E [ESO program 68.B-0310(A)]. 
}
\label{fig8}
\end{figure*}

\begin{figure*}[t]
\hspace*{0.0cm}\psfig{figure=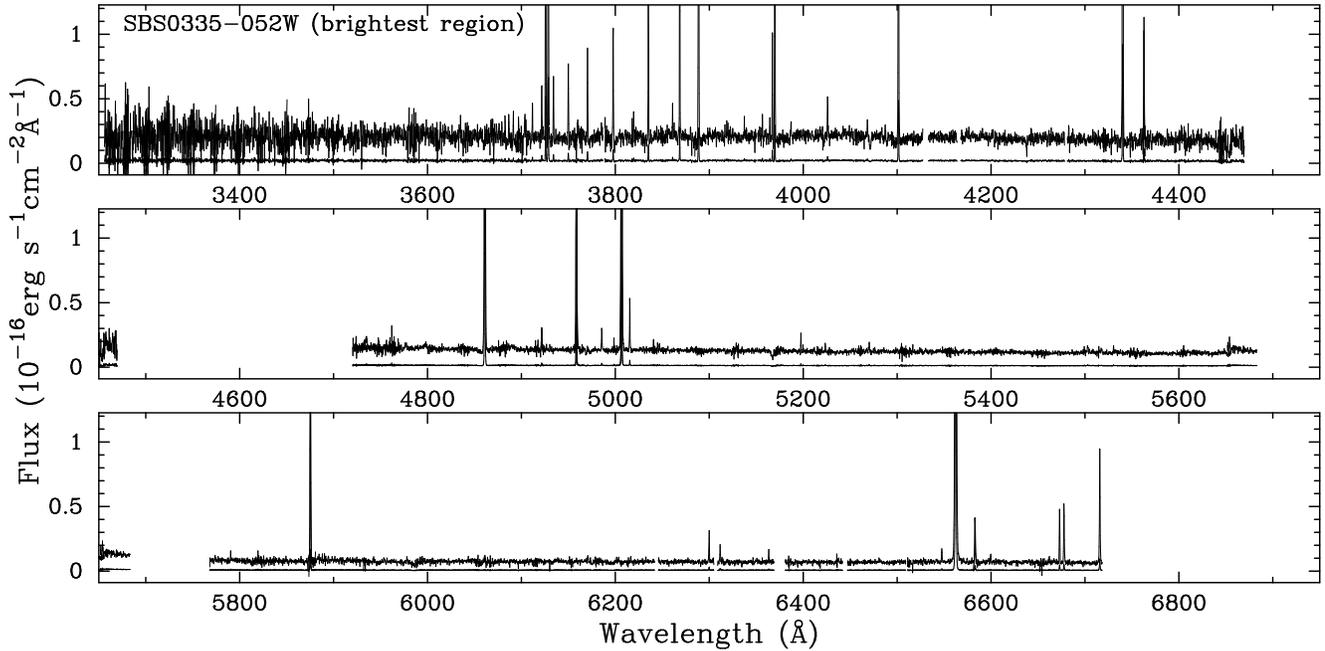,angle=-90,width=17.5cm,clip=}
\caption{Flux-calibrated and redshift-corrected UVES spectrum of 
 SBS 0335--052W [ESO program  70.B-0717(A)]. 
}
\label{fig9}
\end{figure*}

\begin{figure}[t]
\hspace*{0.0cm}\psfig{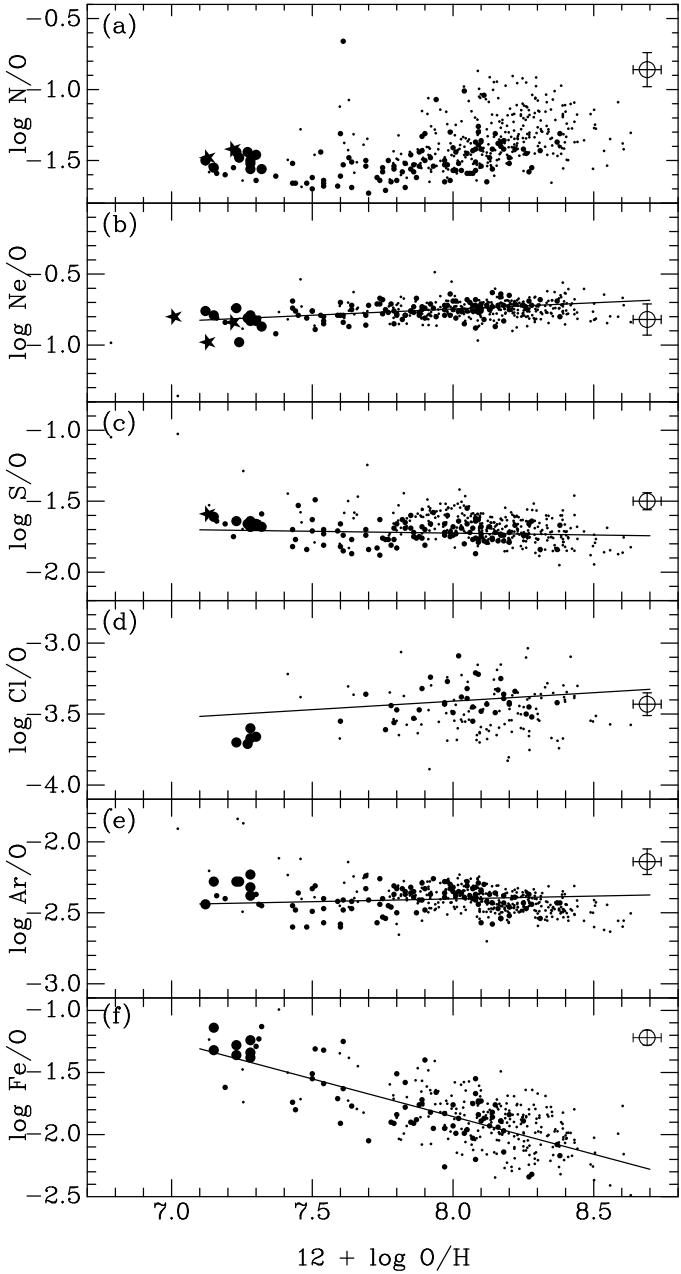}
\caption{log N/O (a), log Ne/O (b), log S/O (c), log Cl/O (d),
log Ar/O (e) and log Fe/O (f) vs oxygen abundance
12 + log O/H for the emission-line galaxies. 
Large filled circles show regions in SBS 0335--052E, large stars 
show regions in SBS 0335--052W. The galaxies from comparison samples are 
shown by small symbols. Small filled circles are galaxies from the
HeBCD sample collected by \citet{ISGT2004} and \citet{IT04a} for the 
primordial He abundance determination,
dots are the galaxies from the SDSS DR3 sample 
\citep{Iz06}.
The solar rations as compiled by \citet{Lodd2003}  
are indicated by the large open circles and the associated error 
bars are shown.
}
\label{fig10}
\end{figure}

The pair of extremely metal-deficient blue compact dwarf (BCD) galaxies
SBS 0335--052E and SBS0335--052W plays a key
role in understanding physical conditions in the interstellar medium, 
star formation, and stellar evolution at very low metallicity. 
The first spectroscopic observations
and abundance determinations of the brighter eastern galaxy were completed 
$\sim$ 20 years ago \citep{I90,I90b,I90c}. The fainter western galaxy was 
discovered and spectroscopically studied first by \citet{P97}. Based on
the VLA H {\sc i} mapping \citet{P01} 
showed  that both galaxies are embedded in a 
large H {\sc i}
cloud (66 by 22 kpc) and constitute a pair (separated by 22 kpc) 
of physically related galaxies.

First abundance determinations of the brightest knot of SBS~0335--052E
inferred a very low oxygen abundance in the range 12+logO/H = 7.0 -- 7.3 
\citep{I90,I90b,I90c,M92}, implying that it is one of the most metal-deficient 
BCD known.
Later spectroscopic studies of SBS 0335--052E indicated that the oxygen
abundances of the brightest regions are 12+logO/H $\sim$ 7.3 
\citep{I97b,I_GIRAF_06,Pap2006}. They also detected variations 
in the oxygen abundance in the range of 12+logO/H = 7.1 -- 7.3 on spatial
scales of $\sim$ 1 -- 2 kpc. On the other hand, the oxygen abundance
in SBS 0335--052W is lower yet, 12+logO/H = 7.12 $\pm$ 0.03 \citep{I05a}, 
making it the lowest-metallicity emission-line galaxy known to date. 
For comparison,
the oxygen abundances in the second and third most metal-deficient galaxies
DDO 68 ($\equiv$SDSS J0956+2849) and I Zw 18 are respectively, 7.14 $\pm$ 0.03
\citep{IT2007} and 7.17 $\pm$ 0.01 \citep{TI2005}.

The extremely low oxygen abundances in the galaxy pair SBS 0335--052E and 
SBS 0335--052W imply that they are among the most suitable relatively nearby 
objects 
(at a distance of 54 Mpc) for studying the properties of low-metallicity
stars and the interstellar medium and 
interpreting the physical conditions in high-redshift
young galaxies. The brighter BCD SBS 0335--052E has been extensively studied
in different wavelength ranges. The most imortant results from these studies
are the following: 1) Star formation in this galaxy resides in several
very compact and bright super-star clusters (SSCs)
\citep{T97,P98,Thompson06,Thompson08}. These SSCs are labelled in 
Fig. \ref{fig1} (right panel) following \citet{T97} and \citet{P98}. 
2) \citet{TI97} discovered evidence of stellar winds in the massive stars 
of SBS 0335--052E based on the {\sl HST} UV spectroscopic observations
of the brightest SSCs. 3) \citet{V00} first detected near-infrared 
emission lines of molecular hydrogen, implying that star formation in 
SBS 0335--052E occurs in molecular clouds. 
\citet{Thompson06,Thompson08} identified this emission with SSCs No.1 and 2.
The discovery of H$_2$ emission is in line with the detection
of warm and hot dust emission \citep{T99,V00, Hunt01,Houck04} that is also 
associated with dense star-forming regions. 4) High-ionization emission lines 
are present in SBS 0335--052E (in particular,
[Ne {\sc v}] $\lambda$3346, 3426) indicating that the hard ionizing radiation 
with photon energies above
7 Ryd is intense in this galaxy. This radiation is most likely 
to be associated with
fast radiative shocks propagating in the dense interstellar medium
\citep{ICS01,TI2005}.

SBS 0335--052W has not been studied in such detail mainly because it is much
fainter than SBS 0335--052E. No {\sl HST} images have yet been obtained 
for this
galaxy precluding detailed analysis of the morphology of its central region.
Ground-based photometric and spectroscopic observations suggest that star 
formation in SBS 0335--052W is confined to several young clusters, which are
not as bright as SSCs in SBS 0335--052E 
\citep{P97,P98,Lip1999,I05a,Pap2006}.
On the other hand, the western galaxy is brighter in the X-ray range 
\citep{T04} and in the H {\sc i} $\lambda$21 cm emission line \citep{P01}.

   Despite extensive optical spectroscopic studies of both galaxies, 
additional high-quality spectroscopic observations with large
telescopes had been required.
First, improving the heavy element abundance determinations 
is important to the study of abundance variations in extremely 
low-metallicity environments 
and understanding the mixing processes in the interstellar medium.
Secondly, this system of galaxies is one of the most suitable targets for the 
determination of the primordial He abundance \citep{ITS07}.
To address these problems, we use 8.2m Very Large Telescope (VLT) 
spectroscopic observations
of both galaxies obtained with the spectrographs FORS1 and UVES
Some of these data are of the highest signal-to-noise ratio
(S/N $\ga$ 100 in the continuum), spatial resolution ($<$1\arcsec), and 
spectral resolution ($\sim$ 0.02 -- 3${\rm \AA}$) ever obtained for these galaxies. 
The raw data
of these observations were extracted from the European Southern Observatory
(ESO) archive. 

The objective of this paper is to determine with the highest
precision, the element abundances in the different regions of
SBS 0335--052E and SBS 0335--052W (Fig. \ref{fig1}), to study heavy element 
abundance variations and derive He abundances.

   In Sect. 2, we describe observations and data reduction, and in Sect. 3 we 
present our results. Our conclusions are presented in Sect. 4.

\section{Observations and data reduction}

\subsection{FORS observations}

The FORS spectra of SBS 0335--052E and SBS 0335--052W were 
obtained on 10 September, 2002 with the FORS1 spectrograph mounted
at the UT3 of the 8.2m ESO VLT [ESO program 69.C-0203(A)].
The observing conditions were photometric throughout the night.

Two sets of spectra were obtained. Low-resolution spectra were
obtained with a grism 300V ($\lambda$$\lambda$$\sim$3850--7500)
and a blocking filter GG 375.
The grisms 600B ($\lambda$$\lambda$$\sim$3560--5970) 
and 600R ($\lambda$$\lambda$$\sim$5330--7480) for the blue 
and red wavelength ranges were used in the high-resolution observations. To 
avoid second-order contamination, the red part of the spectrum was obtained
with the blocking filter GG 435.

  The long ($\sim$418\arcsec) slit with a width of 0\farcs51
was centered on regions No. 1 and 2 in SBS 0335--052E and 
oriented in a direction with position angle of --91\fdg 7,
simultaneously crossing the brightest  H {\sc ii} region in 
SBS 0335--052W (Fig. \ref{fig1}, left).
The spatial scale along the slit was 0\farcs2 pixel$^{-1}$
and the resolving power $\lambda$/$\Delta$ $\lambda$ = 300
in the low-resolution mode and
$\lambda$/$\Delta$ $\lambda$ = 780 and 1160
in a high resolution mode for the 600B and 600R grisms, respectively.
The spectra were obtained at low airmass $\la$ 1.1, 
so no correction for atmospheric refraction was necessary.
The seeing was $\sim$ 0\farcs9 during the low-resolution observations,
0\farcs62 -- 0\farcs77 
during the high-resolution observations in the blue range, and 
1\farcs00 -- 1\farcs07 
during the high-resolution observations in the red range.
The total integration time for the low-resolution observations was 720 s
(6 $\times$ 120 s).
The longer exposures were taken for the high-resolution observations 
and consisted of 
4320 s (6~$\times$~720 s) and 2773 s (4~$\times$~600 s and 1~$\times$~373 s) 
for the blue and red parts, respectively. 
In Fig.~\ref{fig2}, we show the distribution of the 
H$\beta$ emission line flux along the slit. 
The bright regions No. 1+2 and the
much fainter region No. 7 in SBS 0335--052E are labelled in Fig. \ref{fig2}a. 
  The regions No. 1 
through No. 4   in SBS 0335--052W   are denoted in Fig. \ref{fig2}b
according to their brightness in the continuum.
We note, however, that the H$\beta$ line of the region No. 4 is slightly 
stronger than that of the region No. 3.
We note also, that, despite the small angular separation between regions 
No. 2 and No. 3 and between No. 1 and No. 4, the good photometric conditions 
during observations and the small CCD pixel size allowed us to resolve all 
four H {\sc ii} regions in SBS 0335--052W and extract one-dimensional 
spectra for all of them.

\subsection{UVES observations}

  Spectra of regions No. 1+2 of SBS 0335--052E were also obtained on 
6 September 2003 with the UVES echelle spectrograph 
mounted at the VLT (UT2) ESO telescope 
[ESO program 
71.B-0055(A)]. The gratings CD\#2 with the central wavelengths 3900$\AA$
and 4370$\AA$ in the blue arm, 
CD\#3 with the central wavelength 6000$\AA$ and filter BK7-5, and CD\#4 
with the central wavelength 9000$\AA$ and 
filter OG590 in the red arm were used to provide spectra for the
wavelength range $\lambda$$\lambda$3300--10000 over 124 orders.
The slits with length of 8\arcsec\ and 12\arcsec\ for blue and red parts of
the spectra, respectively,  and width 3\arcsec\ were centered on 
regions No. 1+2. The spatial scale along the slit was 0\farcs246 and 
0\farcs182 for the blue and red arms, respectively.
The galaxy was observed at a low average airmass of 1.13, 
so no correction for atmospheric refraction was necessary.
The seeing was 0\farcs61 -- 0\farcs63 at the start and at the end
of observation.
The spectra were obtained at position angle of 0\degr\ with an exposure 
time of 3000 s for the gratings centered on 3900$\AA$ and 6000$\AA$ and of 
6300 s for the gratings centered on 4370$\AA$ and 9000$\AA$.

 Two observations were obtained with UVES on 
9 November  2002 [ESO program 70.B-0717(A)]. 
The first observation was for the brightest part of SBS 0335--052W 
with an exposure time of 17550 s at position angle of 110\degr\  
and the second one simultaneously for regions No. 1+2 and  No. 4+5 of 
SBS 0335--052E with an exposure time of 2400 s at position angle 
of 150\degr\ . 
The grating CD\#2 centered on 3900$\AA$ and filter HER-5 and grating
CD\#3 centered on 5800$\AA$ and filter SHP700 were used resulting in
the wavelength range $\lambda$$\lambda$3250--6950 over 77 orders
for regions No. 4+5 of 
SBS 0335--052E and for the brightest region of SBS 0335--052W).
The slits with length of 8\arcsec\ (grating CD\#2) and 12\arcsec\ 
(grating CD\#3) and width 1\arcsec\ were used.
 The spatial scales along the slit were 0\farcs246 and 0\farcs182 for 
gratings CD\#2 and CD\#3, respectively.
The spectra were obtained at low average airmasses of 1.17 and 1.08 
for SBS 0335--052W and SBS 0335--052E, respectively, 
so no corrections for atmospheric refraction was required.
The seeing was $\sim$ 0\farcs8 during the observations of both galaxies.

 The UVES observations of the regions No. 4+5 and No. 7 of 
SBS 0335--052E were obtained on 11 October  2001 [ESO program 
68.B-0310(A)]. The gratings CD\#1 centered on the wavelength
3460$\AA$ and filter CUS04, and CD\#2 centered on the wavelength
4370$\AA$ and filter CUSO4 were used in the blue arm, while
gratings CD\#3 centered at the wavelength 5800$\AA$ and filter
SHP700 and CD\#4 centered at the wavelength 8600$\AA$ and filter OG590 
were used in the red arm. This setup resulted in the wavelength range 
of $\lambda$$\lambda$3150--9900$\AA$ over 124 orders.
The slit length and width were 12\arcsec\ and 1\farcs5, respectively.
Spatial scales along the slit were 0\farcs246 and 0\farcs182 for 
blue and red arm observations, respectively.
All spectra were obtained at position angle of 60\degr\ with an exposure time 
of 1500 s. Averaged airmasses for CD\#1+CD\#3 and CD\#2+CD\#4 observations
were 1.88 and 2.24. 
Therefore, correction for the atmospheric refraction was 
required, but it was not applied, presumably because the parallactic angle 
of 74\degr\ during observations was close to the position angle. This probably 
has a small effect of atmospheric refraction.
The seeing was 0\farcs77 and 1\farcs26 
at the start and the end of observation, respectively.

\subsection{Data reduction}

The data were reduced with the IRAF\footnote{IRAF is 
the Image Reduction and Analysis Facility distributed by the 
National Optical Astronomy Observatory, which is operated by the 
Association of Universities for Research in Astronomy (AURA) under 
cooperative agreement with the National Science Foundation (NSF).}
software package. This included  bias-subtraction, 
flat-field correction, cosmic-ray removal, wavelength calibration, 
night sky background subtraction, correction for atmospheric extinction, and 
absolute flux calibration of the two-dimensional spectrum.
The spectra were also corrected for interstellar extinction using the 
reddening curve of \citet{W58}.
   The flux-calibrated and redshift-corrected one-dimensional FORS spectra 
of the regions No. 1+2  and 
No. 7 in SBS 0335--052E and four regions
in SBS 0335--052W are shown in Figs.~\ref{fig3} and \ref{fig4},
respectively. 
In Fig. \ref{fig5}, we show the expanded high-resolution
spectrum of SBS 0335--052E No. 1+2 (the same as in Fig. \ref{fig3}) 
to be able to identify more clearly the numerous weak permitted and 
forbidden lines. We note the presence of
broad H$\alpha$, H$\beta$, and H$\gamma$ emission in Fig. \ref{fig5} with
fluxes of $\sim$ 1 -- 2 percent of the narrow component fluxes, in
agreement with the value for the H$\beta$ emission line 
obtained by \citet{ITG07} for the same regions No. 1+2.
   The flux-calibrated and redshift-corrected UVES spectra of regions No. 1+2,
No. 4+5, and  
No. 7 of SBS 0335--052E are shown in Figs.~\ref{fig6},~\ref{fig7}, and 
\ref{fig8}, respectively, and the spectrum of the bright part of 
SBS 0335--052W is shown in Fig.~\ref{fig9}.

Emission-line fluxes were measured using Gaussian profile fitting. 
The 1$\sigma$ errors of the line fluxes were calculated from the photon 
statistics in the non-flux-calibrated spectra. 
The true errors in the line fluxes are
probably higher because we do not take into account uncertainties introduced
during observations (e.g., effect of differential refraction) and data
reduction. The line flux errors were propagated in the 
calculations of the elemental abundance errors.
The extinction-corrected emission line fluxes $I$($\lambda$) relative to the 
H$\beta$ fluxes multiplied by 100, the extinction coefficients 
$C$(H$\beta$), the 
equivalent widths EW(H$\beta$),
the observed H$\beta$ fluxes $F$(H$\beta$) and the 
equivalent widths of the hydrogen absorption lines
are listed in Tables~\ref{tab1} and \ref{tab2} (low-resolution and 
high-resolution FORS observations), in Table~\ref{tab3} (for weak lines 
in the high-resolution FORS spectrum of SBS 0335--052E No. 1+2),
and in Table~\ref{tab4} (UVES observations).

\section{Results}

\subsection{Electron temperature and electron number density}

The electron temperature $T_e$, the 
ionic and total heavy element abundances were derived 
following \citet{Iz06}. In particular for 
O$^{2+}$, Ne$^{2+}$ and Ar$^{3+}$, we adopt
the temperature $T_e$(O {\sc iii}) directly derived from the 
[O {\sc iii}] $\lambda$4363/($\lambda$4959 + $\lambda$5007)
emission-line ratio. We use $T_e$(O {\sc ii}) for the calculation of
O$^{+}$,  N$^{+}$, S$^{+}$, and Fe$^{2+}$ abundances, and $T_e$(S {\sc iii})
for the calculation of S$^{2+}$, Cl$^{2+}$, and Ar$^{2+}$ abundances.

For all regions in SBS 0335--052E, the electron number densities 
$N_e$(S {\sc ii}) were obtained from the [S {\sc ii}] 
$\lambda$6717/$\lambda$6731 emission line ratio, whereas 
in all regions in SBS 0335--052W, except for region No. 1, 
the [S {\sc ii}] $\lambda$6717, 6731 emission lines are too weak 
to allow density determinations.
Therefore, for abundance determinations
in regions No. 2, 3, and 4 of SBS 0335--052W, we adopt $N_e$ = 10 cm$^{-3}$.
The value of the electron number density 
makes little difference to the derived abundances
since in the low-density limit, which holds for the H {\sc ii} regions
considered here, the element abundances do not depend sensitively 
on $N_e$. We note that errors of the element abundances do not account for the 
uncertainties in the ionization correction factors.
The electron temperatures $T_e$(O {\sc iii}), 
$T_e$(O {\sc ii}), and $T_e$(S {\sc iii}), 
electron number density $N_e$(S {\sc ii}), 
the ionization correction factors ($ICF$s), and
the ionic and total N, O, Ne, S, Cl, Ar, and Fe abundances derived from the
forbidden emission lines
are given in Tables \ref{tab5} and \ref{tab6} (low-resolution and
high-resolution FORS observations), respectively and in 
Table~\ref{tab7} (UVES observations). 
We note that the $T_e$(S {\sc iii}) derived in all observations lies between
$T_e$(O {\sc iii}) and $T_e$(O {\sc ii}), as expected.

\subsection{Oxygen abundance}

Abundances of O$^+$ and O$^{2+}$ in all regions of SBS 0335--052E and
SBS 0335--052W were obtained from the fluxes of forbidden
emission lines [O {\sc ii}] $\lambda$3727 and [O {\sc iii}]
$\lambda$4959, 5007. Additionally, for the determination of the
O$^{2+}$ abundance, we used the very weak recombination emission 
line O {\sc ii} $\lambda$4650, which was detected 
in the high-resolution FORS spectrum of regions No. 1+2 in 
SBS 0335--052E (Table \ref{tab4}). 
Adopting radiative recombination coefficients from \citet{P91}, 
we obtained a O$^{2+}$/H$^+$ abundance ratio of 
(1.50$\pm$0.75)$\times$10$^{-5}$ from the recombination line flux. 
This value is consistent with that derived from the forbidden 
lines (Table \ref{tab6}).
   
We derived the oxygen abundances of regions 1+2 in 
SBS 0335--052E to be 12 + log O/H = 7.28$\pm$0.01 (low-resolution FORS
observations), 7.23$\pm$0.01 (high-resolution FORS observations), and
7.28$\pm$0.01 and 7.27$\pm$0.01 (for two different UVES observations).
For regions 4+5, the oxygen abundance was inferred to be 
12 + log O/H = 7.28$\pm$0.01 and 
7.32$\pm$0.01 for two UVES observations. For cluster 7,
its oxygen abundance is 7.12$\pm$0.04 (low-resolution FORS 
observation), 7.15$\pm$0.02 (high-resolution FORS observation), 
and 7.24$\pm$0.05 (UVES observation).
 The true differences in the oxygen abundance determinations for the same
regions are higher than the calculated errors based on the noise statistics,
implying that the errors could be underestimated. In particular, these
errors do not include 
uncertainties introduced by the standard data reduction. Additionally,
differences in oxygen abundances could probably be caused by
different slit positions and different apertures.
The oxygen abundance obtained from a combined spectrum of the 
brightest regions 1+2+4+5 is 7.30$\pm$0.01 and is very close to the value 
7.31$\pm$0.01 obtained by \citet {I97b} and \citet{TI2005} for the brightest 
part of SBS 0335--052E .
Our oxygen abundances for regions 4+5 are close to the value 
7.27$\pm$0.02 obtained by \citet{Pap2006}.
We find that the oxygen abundance in the fainter region 7 is lower
than in the brighter regions and compares well with the previous determination
of 7.21$\pm$0.02 by \citet{Pap2006}.
Thus,  we confirm the tendency for the oxygen abundance to decrease from
the brightest part of SBS 0335--052E to its outer fainter part, 
suggesting the presence of oxygen abundance variations on spatial scales 
of $\sim$ 1 -- 2 kpc, 
in agreement with the finding of \citet{I97b}, \citet{I_GIRAF_06}, 
and \citet{Pap2006}.

The oxygen abundance of the whole bright part of SBS 0335--052W 
obtained from the UVES spectrum is 12 + log O/H = 7.13 $\pm$ 0.02. 
  This value is consistent with  
12 + log O/H = 7.12 $\pm$ 0.03 found by \citet{I05a} from combined 
4m Kitt Peak, 6.5m MMT, and 10m Keck II telescope observations,
with 12 + log O/H = 7.11 $\pm$ 0.05 obtained by \citet{TI2005} 
from MMT observations, and with
12 + log O/H = 7.13 $\pm$ 0.08 found by \citet{Pap2006} from
3.6m ESO telescope observations.
  Thus, different determinations of the oxygen abundance in the aperture 
covering the bright part of SBS 0335--052W are in a good agreement.
  
The determination of the oxygen abundances in individual star-forming regions
of SBS 0335--052W is a more difficult task because of the faintness of these 
regions and the small angular separations between them.
From 3.5m Calar Alto telescope $R$, $I$ photometry, \citet{Lip1999} have found 
that SBS 0335--052W consists at least of three star-forming regions. 
Given the good seeing during FORS observations, we find that SBS 0335--052W
consists of 4 regions (Fig. \ref{fig2}b).
Our regions No. 1 and 2 correspond to the brightest western and eastern regions
of \citet{Lip1999}.
The oxygen abundance of 12 + log O/H = 7.22 $\pm$ 0.07 of the brightest 
region No. 1 (FORS) is consistent with the value
12 + log O/H = 7.22 $\pm$ 0.03 inferred by \citet{Lip1999}. However, our oxygen
abundance of 12 + log O/H = 7.01 $\pm$ 0.07 for region No. 2 is 
significantly lower than that of 12 + log O/H = 7.13 $\pm$ 0.07 obtained 
by \citet{Lip1999}. The main reason for the difference is that \citet{Lip1999}
did not observe the blue part of the spectrum in region No. 2 covering 
the emission line [O {\sc ii}] $\lambda$3727 and assumed its flux relative 
to H$\beta$ to be the same as that in region No. 1. The relative 
flux [O {\sc ii}] $\lambda$3727/H$\beta$ in this region is a factor 
of $\sim$2 lower than that adopted by \citet{Lip1999} (see Table \ref{tab2}). 
Applying our value of [O {\sc ii}] $\lambda$3727/H$\beta$ ratio to the
\citet{Lip1999} observations, we obtain 12 + log O/H = 7.04 $\pm$ 0.07, which
is very close to the value for region No. 2 obtained from our data.
  The fainter regions No. 3 and 4 in SBS 0335--052W were not discussed in 
previous papers and for the first time, we derive element abundances
in these regions. 
Thus, we confirm the very low oxygen abundance of region 1.
We also find that three of four H {\sc ii} regions have 
unprecedently low oxygen abundances of 12 + log O/H = 7.01 $\pm$ 0.07 
(region 2), 6.98 $\pm$ 0.06 (region 3), and 6.86 $\pm$ 0.14 (region 4),
confirming our previous findings that this galaxy is the most 
metal-deficient emission-line galaxy known. 
   We note, that [O {\sc iii}] $\lambda$4363 emission line was not
detected for region 3. Therefore, the oxygen abundance in this region
was derived using the semi-empirical method described by \citet{IT2007}.
Similar to SBS 0335--052E, we find that the oxygen abundance in 
SBS 0335--052W varies from region to region. These variations are in the range 
from 12 + log O/H = 6.87 to 7.22, which is even higher than the variations 
that we obtained for SBS 0335--052E.

If real, the oxygen abundance variations in both galaxies would argue
in favour of self-enrichment by the fresh heavy elements synthesized during
the present burst of star formation and slow mixing of these elements with
the surrounding regions. It is also possible that the 
[O {\sc iii}] $\lambda$4363$\AA$ emission line is enhanced by shocks and
the effect is likely to be stronger in regions with weaker lines
\citep{Pe91,ITL97}. Consequently,
this enhancement would result in the underestimation of the oxygen abundance.
However, this effect is difficult to take into account because of insufficient
information about the shock contribution to ionization and heating of
H {\sc ii} regions.
  
\subsection{Other heavy element abundances}

  In Fig.~\ref{fig10}, we show the abundance ratios  log N/O (a), log Ne/O (b),
log S/O (c), log Cl/O (d), log Ar/O (e), and log Fe/O (f) versus 
oxygen abundance
12 + log O/H for different H {\sc ii} regions in SBS 0335--052E and 
SBS 0335--052W,
and compare them with data for a large sample of emission-line galaxies. 
Data for H {\sc ii} regions in SBS 0335--052E are shown by large filled 
circles and for H {\sc ii} regions in SBS 0335--052W by stars. 
The galaxies from comparison samples are 
shown by small symbols. We show by small filled circles data for galaxies
collected to study the helium abundances 
in low-metallicity blue compact dwarf galaxies 
\citep[the HeBCD sample,][]{ISGT2004,IT04a}, and by
dots the galaxies from the SDSS DR3 sample \citep{Iz06}.
In each panel, the solar abundance ratio by \citet{Lodd2003}  
is indicated by the large open circle and the associated error bar.
 
As for $\alpha$-elements, the ratios of Ne, S, Ar abundances to oxygen 
abundance, Ne/O
derived for different regions in SBS 0335--052E and SBS 0335--052W
follow the trend of increasing Ne/O with oxygen abundance
found by \citet{Iz06} for other low-metallicity emission-line galaxies
(Fig. \ref{fig10}b).
This trend is caused by oxygen depletion onto dust and implies that there is
a small amount of
depletion of oxygen in SBS 0335--052E and SBS 0335--052W despite
the detection of dust in SBS 0335--052E \citep[e.g., ][]{T99,Houck04}.

The S/O abundance ratio in both galaxies is close to the average value 
obtained for other
low-metallicity emission-line galaxies (Fig. \ref{fig10}c). On the other
hand, Cl/O in SBS 0335--052E is systematically lower (Fig. \ref{fig10}d)
and Ar/O is systematically higher (Fig. \ref{fig10}e) than the averaged 
values obtained by \citet{Iz06} for other low-metallicity galaxies.
These differences imply that there is a metallicity dependence 
of the $\alpha$-element production by massive stars. 
The explosion energy of Type II supernovae might play a role
\citep{K06}. However, \citet{Iz06}
pointed out that abundance determinations for some elements such as S, Cl, and
Ar could be uncertain because of the uncertainties in the rates of some
atomic processes, e.g., rates of dielectronic recombination. Further analysis
of larger samples of extremely low-metallicity galaxies is needed to
clarify the reasons for Cl/O and Ar/O deviations in SBS 0335--052E from
other galaxies.

The most prominent trend was found by \citet{Iz06} for the Fe/O ratio.
New data for SBS 0335--052E confirm and
strengthen the previous result that Fe is depleted onto dust grains
and that this effect depends on the galaxy metallicity.
The depletion of Fe decreases with decreasing metallicity, and in the 
H {\sc ii} regions of SBS 0335--052E with the lowest 
metallicity, the depletion is the lowest. 

  \citet{IT99} and \citet{Iz06} demonstrated that the dispersion in N/O in low 
metallicity BCDs with 12 + log O/H $<$ 7.5 -- 7.6 is very small with a 
plateau value of log N/O $\sim$ --1.6. However, in this paper, we find 
that N/O in H {\sc ii} regions of SBS 0335--052E and SBS 0335--052W
is higher than a plateau value implying that there is some increase 
in N/O with decreasing oxygen abundance at 12 + log O/H $<$ 7.5
(Fig. \ref{fig10}a).
If true, this tendency would agree with studies of 
primary nitrogen production by low-metallicity rotating stars. 
For example, \cite{MM02} showed that the production of primary nitrogen
is significantly higher in rapidly rotating stars of extremely low metallicity
as compared to that in non-rotating stars.
Although they considered stellar models with the heavy element mass fraction
$Z$ = 10$^{-5}$, which is significantly lower than the values of 
$Z$ $\sim$ 0.0002 -- 0.0005 in
the H {\sc ii} regions of SBS 0335--052E and SBS 0335--052W, it is 
possible that the interstellar medium in these galaxies
``memorizes'' the chemical enrichment by the most metal-deficient 
stars.

\begin{figure*}[t]
\hspace*{0.0cm}\psfig{figure=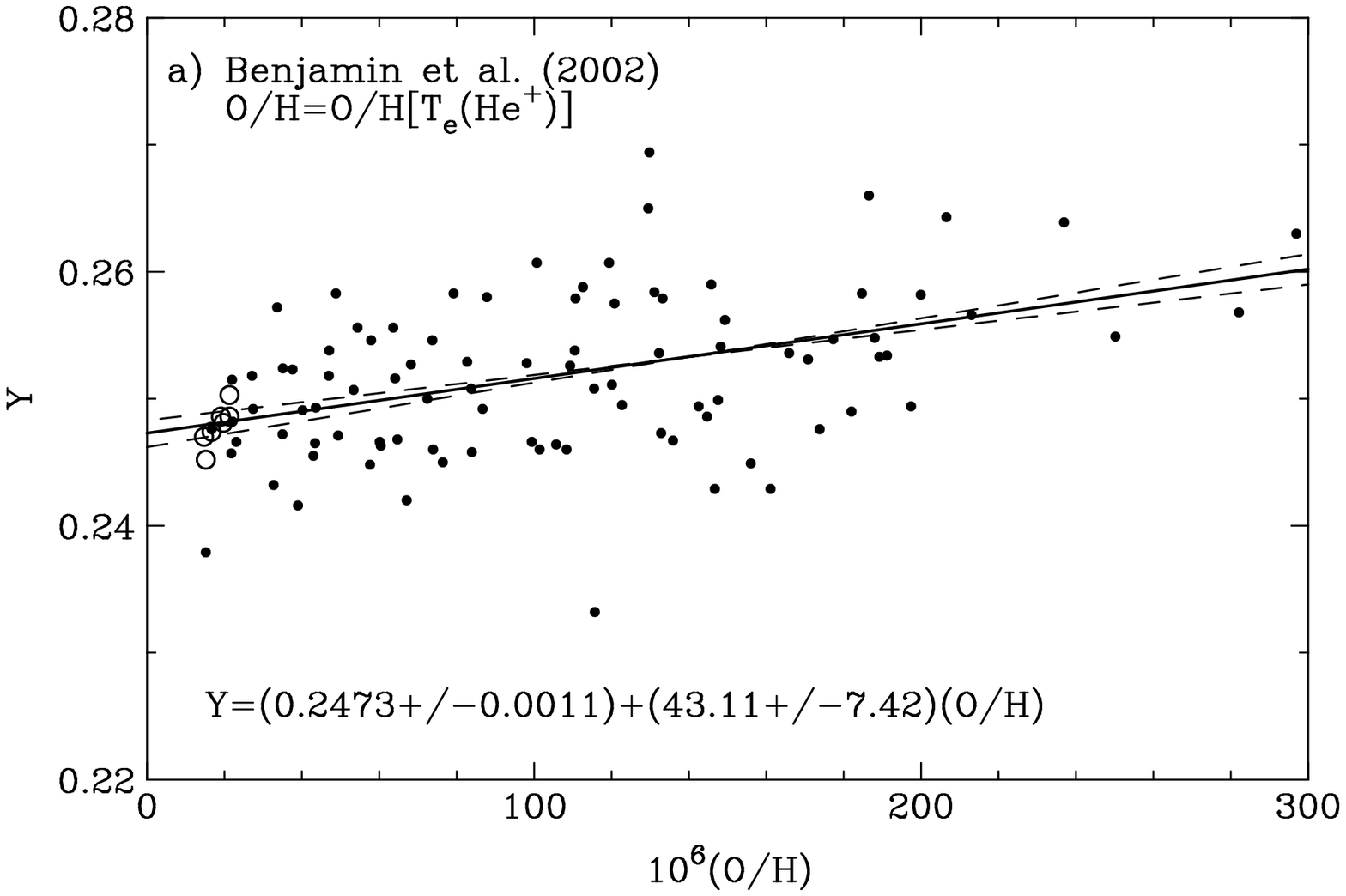,angle=0,width=9.cm,clip=}
\hspace*{0.0cm}\psfig{figure=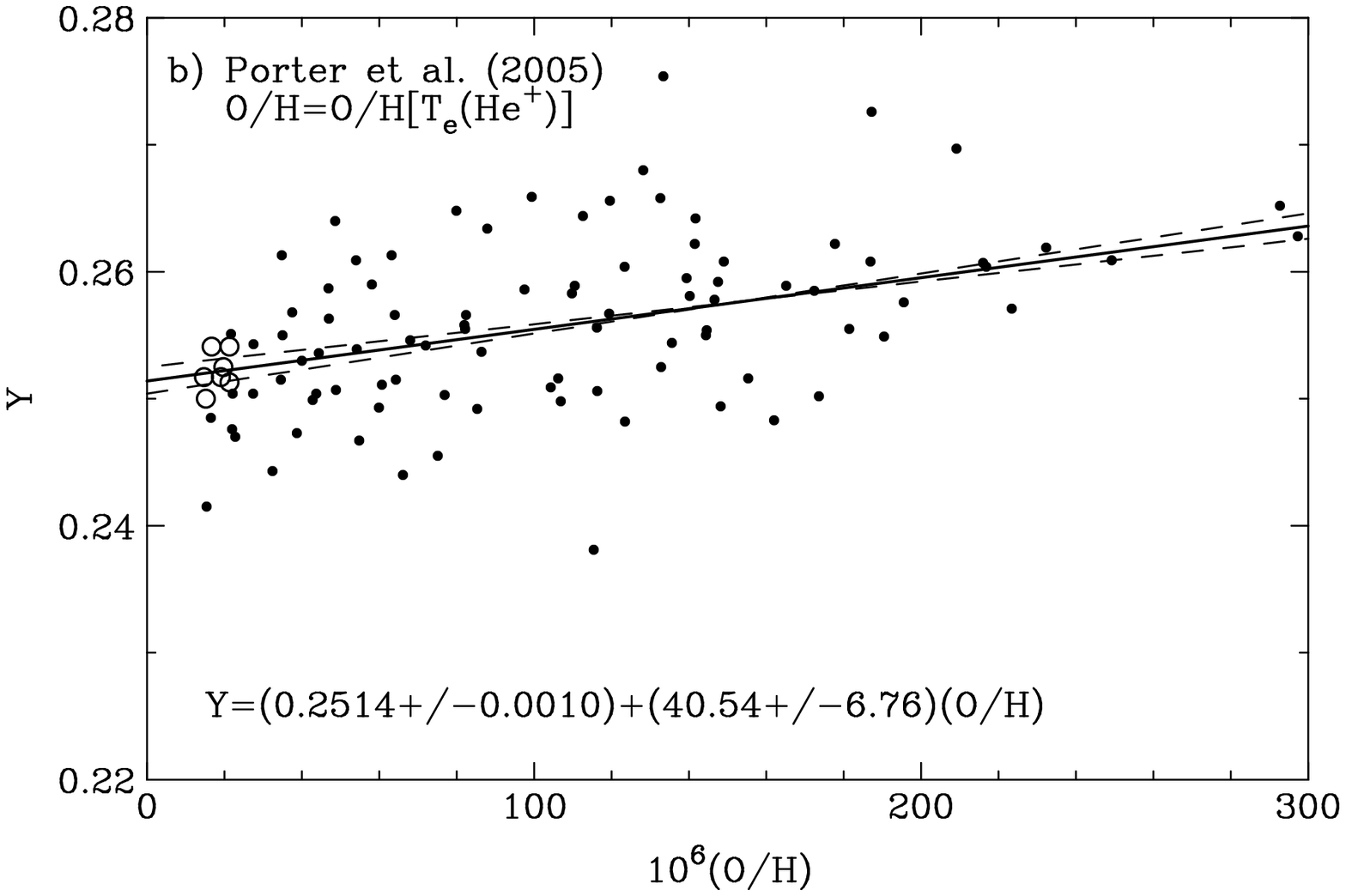,angle=0,width=9.cm,clip=}
\caption{Linear regressions of the helium mass fraction $Y$ vs. oxygen 
abundance for the sample consisting of H {\sc ii} regions in SBS 0335--052E
and SBS 0335--052W (open circles, this paper), and HeBCD H {\sc ii}
regions (dots) from \citet{ITS07}. Solid lines are linear regressions
obtained by a likelihood method, and dashed lines are their 1$\sigma$ 
alternatives.
The $Y$s in (a) and (b) are derived with the He {\sc i} emissivities
from \citet{B99,B02} and \citet{Po05}, respectively. 
The oxygen abundance is derived adopting $T_e$(He$^+$).
}
\label{fig11}
\end{figure*}

\subsection{Helium abundance}

Helium ($^4$He) and some other light elements and isotopes 
such as D, $^3$He, and $^7$Li were produced mainly within the first
few minutes after the Big Bang and they can therefore be used
to constrain cosmological models \citep[e.g., ][]{S05,S06}.
Because of their extremely low metallicity, SBS 0335--052E and SBS 0335--052W
are among the most suitable objects to use for the primordial 
He abundance determination. 
This is because the fraction of He produced by stars in these galaxies is
less than 1\% of the fraction of He synthesized during the primordial
nucleosynthesis. Significant efforts have been made to determine
the primordial He mass fraction $Y_p$ from observations of 
low-metallicity H {\sc ii} regions and the baryonic mass fraction of the
Universe $\Omega$$_b$. 
Because of the logarithmic dependence of $Y_p$ on $\Omega$$_b$, 
an accuracy of higher than 1\% in $Y_p$ is required to place 
constraints on the cosmological models. This high accuracy is feasible
for the data obtained with the VLT for SBS 0335--052E and SBS 0335--052W
\citep[provided that the atomic data are known with sufficient accuracy, ][]{P09}.
In this paper, we determine the He abundance following
the method described by \citet{ITS07}.

There is now a general 
consensus that the accuracy of the  primordial He abundance determination
is limited, not so much 
by statistical uncertainties, but by our ability to account for systematic 
errors and biases \citep[e.g., ][]{OS04,P07,P09}.   
There are many known effects to correct for when  
transforming observed He {\sc i} line intensities into a He abundance. 
These effects are: (1) reddening; (2) underlying stellar
absorption in the He {\sc i} lines; (3) collisional 
excitation of the He {\sc i} lines, which causes their intensities to
 deviate from their recombination values; 
(4) fluorescence of the He {\sc i} lines, which also make their intensities
deviate from their 
recombination values; (5) collisional excitation of the hydrogen lines 
(hydrogen is relevant because the helium abundance is calculated relative to 
that of hydrogen); 
(6) possible departures from case B in the emissivities of H and He {\sc i} 
lines\footnote{Case B assumes that there exists a balance between the 
absorption and emission of photons in the resonant Lyman series transitions of 
hydrogen and helium and that there are no other processes.}; 
(7) the temperature 
structure of the H {\sc ii} region; and (8) its ionization structure. All of
these corrections are at a level of a few percent 
except for effect (3), which may be much higher, exceeding 10\%, in the case
of the He {\sc i} $\lambda$5876 emission line in hot 
and dense H {\sc ii} regions. 

The derived He$^+$ abundance $y^+$ = He$^+$/H$^+$ depends on the adopted 
He {\sc i} line emissivities. 
We considered two sets of He {\sc i} emissivities: older values by 
\citet{B99,B02}, 
which were used by \citet{IT04a} [\citet{B02} take into account both 
collisional and fluorescent 
enhancements], and more recent values by \citet{Po05}, 
which were computed using improved radiative and collisional data. 
Following \citet{ITL94,ITL97} and \citet{IT98}, we used the five 
strongest He {\sc i} $\lambda$3889, $\lambda$4471, 
$\lambda$5876, $\lambda$6678, and $\lambda$7065 emission lines to derive
the weighted mean He$^+$ abundance $y^+_{wm}$, where the weights were defined 
by the flux error of each He {\sc i} emission line. 

In addition to the emissivities,
 the derived $y^+$ abundances also depend on a number of  
other parameters: the fraction $\Delta$$I$(H$\alpha$)/$I$(H$\alpha$) of the
H$\alpha$ emission line flux due to collisional excitation; the electron
number density $N_e$(He$^+$); the electron temperature $T_e$(He$^+$); 
the equivalent
widths EW$_{abs}$($\lambda$3889), EW$_{abs}$($\lambda$4471), 
EW$_{abs}$($\lambda$5876), 
EW$_{abs}$($\lambda$6678), and EW$_{abs}$($\lambda$7065) of He {\sc i} stellar 
absorption lines; and the optical depth 
$\tau$($\lambda$3889) of the He {\sc i} $\lambda$3889 emission line. 
To determine the most robust value of $y^+_{wm}$, 
we use the Monte Carlo procedure described in \citet{IT04a}, 
randomly varying each of the above parameters within a specified range. 
First, we took into account collisional excitation effects for hydrogen. 
The fraction of the H$\alpha$ flux produced by collisional  
excitation was randomly generated 100 times within an 
adopted range. The fraction of
the H$\beta$ emission line flux due to the collisional excitation
was adopted to be three times less than that of the H$\alpha$ flux. 
For each generated fraction, the fluxes of the H$\alpha$ and H$\beta$ lines
caused by the collisional excitation were subtracted from the total observed
fluxes and all emission line fluxes were then corrected for underlying stellar 
absorption (in the case of hydrogen lines) and interstellar
extinction, and element abundances were calculated.

To calculate $y^+$, we varied $N_e$(He$^+$) simultaneously and randomly 
in the range 10 -- 450 cm$^{-3}$,
$T_e$(He$^+$) in the range (0.95 -- 1.0)$\times$$T_e$(O {\sc iii}),
and $\tau$($\lambda$3889) in the range 0 -- 5. We produced a 
total of 10$^5$ of these realizations for every H {\sc ii} region, for
a given fraction of H$\alpha$ emission line flux created by 
collisional excitation.
Thus, the total
number of Monte Carlo realizations that we performed 
for each H {\sc ii} region was 
100 $\times$ 10$^5$ = 10$^7$.
As for the He {\sc i} underlying stellar absorption, we followed
prescriptions by \citet{ITS07}. We adopted fixed values of
EW$_{abs}$($\lambda$4471) = 0.4$\AA$,   
EW$_{abs}$($\lambda$3889) / EW$_{abs}$($\lambda$4471) = 1.0,
EW$_{abs}$($\lambda$5876) / EW$_{abs}$($\lambda$4471) = 0.3,
EW$_{abs}$($\lambda$6678) / EW$_{abs}$($\lambda$4471) = 0.1,
and EW$_{abs}$($\lambda$7065) / EW$_{abs}$($\lambda$4471) = 0.1.

For each H {\sc ii} region, we found the optimal solution for $y^+_{wm}$ in 
the multi-parameter space defined above by 
minimizing the quantity
\begin{equation}
\chi^2=\sum_i^n\frac{(y^+_i-y^+_{wm})^2}{\sigma^2(y^+_i)}\label{eq1},
\end{equation}
where $y^+_i$ is the He$^+$ abundance derived from the flux of the He {\sc i}
emission line labelled $i$, and $\sigma(y^+_i)$ is the statistical error
of $y^+_i$. The quantity $y^+_{wm}$ is the weighted 
mean of the He$^+$
abundance given by the equation
\begin{equation}
y^+_{wm}=\frac{\sum_i^k{y^+_i/\sigma^2(y^+_i)}}
{\sum_i^k{1/\sigma^2(y^+_i)}}\label{eq2}.
\end{equation}
We used all five He {\sc i} emission lines to calculate $\chi^2$ (i.e.,
$n$ = 5), but only three lines, He {\sc i} $\lambda$4471,
$\lambda$5876, and $\lambda$6678 to compute
$y^+_{wm}$ ($k$ = 3). This is because the fluxes of the
He {\sc i} $\lambda$3889 and $\lambda$7065 emission lines are
more uncertain than those of the other three He {\sc i} emission lines.

Additionally, in those cases when the nebular He {\sc ii} $\lambda$4686 
emission line was detected, we added the abundance of doubly ionized 
helium $y^{2+}$ $\equiv$ He$^{2+}$/H$^+$ to $y^+$. Although the He$^{2+}$ zone
is hotter than the He$^{+}$ zone, we 
adopted $T_e$(He$^{2+}$) = $T_e$(He$^{+}$).
This assumption has only a minor effect on the $y$ value, because
$y$$^{2+}$ is small ($\leq$ 3\% of $y^+$) in all cases.
Then the total He abundance was derived to be
\begin{equation} 
y = ICF({\rm He}^+ +{\rm He}^{2+}) \times (y^+ + y^{2+}),
\end{equation} 
where
$ICF$(He$^+$+He$^{2+}$) is the ionization correction factor, adopted
from \citet{ITS07}.
The He mass fraction was obtained from
\begin{equation}
Y=\frac{4y(1-Z)}{1+4y}, \label{eq:Y}
\end{equation}
where $Z$ is the heavy element mass fraction. 
We adopted $Z$ = $a$(O/H) 
in Eq. \ref{eq:Y}, where $a$ = 18.2 was obtained by assuming 
an oxygen mass fraction of O=0.66$Z$ 
for a heavy element mass fraction $Z$=0.001 \citep{Ma92}.
Some authors use slightly different values of $a$. In particular,
\citet{P92} adopted $a$ = 20. These differences introduced a tiny effect 
on the $Y$ value in H {\sc ii} regions with O/H $\la$ 10$^{-4}$, resulting in 
a difference of $\la$0.02\% in $Y$, if our value of $a$ is adopted instead of 
that by \citet{P92}.

The results of our determinations of the He abundance in different regions
of SBS 0335--052E and SBS 0335--052W corresponding to minimum of
$\chi^2$ for two different sets of emissivities are shown in Table \ref{tab8}.
It can be seen that the derived He mass fractions $Y$ are very consistent 
(within $\la$ 1\%) in all H {\sc ii} regions,
while the statistical errors 1 -- 5\% of the $Y$ values are comparable to 
those obtained by other authors for different galaxies 
\citep[see e.g., ][]{P07}. 
The weighted mean $Y$ derived
from all observations shown in Table \ref{tab8} is equal to 
$Y_{wm}$ = 0.2485 $\pm$ 0.0012 for the emissivities derived by \citet{B99,B02} 
and $Y_{wm}$ = 0.2514 $\pm$ 0.0012 for \citet{Po05} emissivities. These values
are very close to the primordial He abundance since the considered H {\sc ii}
regions are of very low-metallicity.
They are obtained by taking into account almost all systematic effects. 
The largest
systematic effect is collisional excitation of He {\sc i} emission lines.
In the case of SBS 0335--052E No.1+2 this effect could increase the intensity
of the strongest He {\sc i} $\lambda$5876 emission line by up to 10\% 
relative to the recombination value. Collisional excitation of the He {\sc i} 
$\lambda$7065 emission line has an even stronger effect, of $\sim$50\%. 
If not taken into account, this systematic effect can produce
the weighted mean He mass fraction of 0.266, or a value $\sim$ 6\% higher
than the value of 0.251 in Table \ref{tab8} for \citet{Po05} emissivities.
This difference is much higher than the errors of $\sim$ 1-3\% for $Y_p$ 
obtained by different authors. However, since these systematic effects are
taken into account in our $Y$ determination, they are not included in the
error budget. An exception is the differences between two different sets
of He {\sc i} emissivities by \citet{B99,B02} and \citet{Po05}. 
Treating the different 
He {\sc i} line emissivities as systematics affecting the $Y_{wm}$ 
determination, we obtain
$Y_{wm}$ = 0.2500 $\pm$ 0.0012 (stat.) $\pm$ 0.0015 (syst.).

We also rederive the primordial He abundance from linear regressions 
to $Y$ -- O/H for the large sample of low-metallicity H {\sc ii} regions from 
\citet{ITS07}, including also our values 
for SBS 0335--052E and SBS 0335--052W from the present study. 
These regressions are obtained by taking into account the statistical errors 
in $Y$ and O/H for every object and are shown
in Fig. \ref{fig11} for two sets of emissivities, \citet{B99,B02} 
(Fig. \ref{fig11}a) and \citet{Po05} (Fig. \ref{fig11}b). In this Figure, we
indicate by dots 93 observations of 77 H {\sc ii} regions
from the HeBCD sample \citep{IT04a,ITS07} and 
by open circles different H {\sc ii} regions in SBS 0335--052E and 
SBS 0335--052W.
Solid lines are linear regressions obtained by a likelihood method, and the 
dashed lines are 1$\sigma$ alternatives to these regressions. 
We obtain $Y_p$ = 0.2473 $\pm$ 0.0011 for the \citet{B99,B02} emissivities and
$Y_p$ = 0.2514 $\pm$ 0.0010 for the \citet{Po05} emissivities. 
Despite the statistical errors of $Y$ for the H {\sc ii} 
regions in our sample being comparable to those derived in other studies, the 
statistical errors of $Y_p$ are significantly lower than those derived by 
e.g., \citet{OS04}, \citet{FK06}, and \citet{P07}. 
This difference is present because 
we use the largest sample of $\sim$ 100 H {\sc ii} regions to
determine $Y_p$ compared to 30 or even only 6 H {\sc ii} regions
in other studies. The error in $Y_p$ scales as $\sim$ 1/$\sqrt{N}$,
where $N$ is the number of the objects used for the primordial He abundance
determination, resulting in the reduction in the $Y_p$ error by a factor 
of 2 -- 4 for our large sample. Our
values of $Y_p$ are only slightly below ($\la$ 1\%) the values of $Y_{wm}$
obtained for the BCD system SBS 0335--052 with the respective 
emissivities, but they are consistent within the errors.
The value of $Y_p$ obtained with the \citet{B99,B02}
emissivities is in good agreement with the value of $Y_p$ =
0.2482$^{+0.0003}_{-0.0004}$ $\pm$ 0.0006 (syst) inferred by 
\citet{S07} from the WMAP data analysis. On the other hand, the $Y_p$
derived with the \citet{Po05} emissivities is slightly above the
WMAP value. If real, this higher value would suggest some deviations
from the Standard Big Bang Nucleosynthesis model.
On the other hand, treating the different He {\sc i} line emissivities as 
systematics affecting the $Y_p$ determination, we obtain
$Y_p$ = 0.2493 $\pm$ 0.0011 (stat.) $\pm$ 0.0020 (syst.), which agrees
with the $Y_p$ derived from the WMAP data analysis. 

\section{Summary}

    We have presented our analysis of archival VLT/FORS and VLT/UVES 
spectroscopic
observations of a system of two extremely low-metallicity blue
compact dwarf (BCD) galaxies, SBS 0335--052E and SBS 0335--052W.
Some of these spectroscopic data are of the highest signal-to-noise ratio,
highest spectral resolution and highest spatial resolution obtained
up to date for this system of galaxies since its first spectroscopic
observations, 20 years ago \citep{I90,I90b,I90c}.

Our main results are as follows:

1. The oxygen abundance in different regions of both 
galaxies is extremely low. We find that the oxygen abundance
12 + log O/H in SBS 0335--052E varies in the range 7.12 -- 7.32,
while in SBS 0335--052W, it varies in the range 6.86 - 7.22.
Thus, SBS 0335--052W holds the record as the most metal-deficient
emission-line galaxy known, supporting and strengthening previous
findings by \citet{I05a} and \citet{Pap2006}. 
We suggest that the oxygen abundance variations in both galaxies
are likely to be real and imply incomplete mixing of the interstellar
medium and the chemical element self-enrichment of H {\sc ii} regions by 
the present bursts of star formation.

2. We derive the He mass fraction $Y$ in different H {\sc ii} regions
of SBS 0335--052E and SBS 0335--052W using
Monte Carlo simulations and taking into account systematic
effects. We find that the He mass fraction in all studied H {\sc ii} regions
varies only a little and, depending on the adopted He {\sc i} line
emissivities, has a weighted mean 
value $Y_{wm}$ of 0.2485$\pm$0.0012
with the \citet{B99,B02} emissivities and of 0.2514$\pm$0.0012 with the
\citet{Po05} emissivities. 
Using the data of the present paper for
SBS 0335--052E and SBS 0335--052W in connection with 
the HeBCD sample of low-metallicity
BCDs by \citet{IT04a} and 
\citet{ITS07}, we rederive the primordial He mass fraction $Y_p$ 
from linear regressions to $Y$ versus O/H. 
We determine a value of 
$Y_p$ = 0.2473$\pm$0.0011 using the \citet{B99,B02} He {\sc i}
emissivities and of 0.2514$\pm$0.0010 using the 
\citet{Po05} He {\sc i} emissivities. 
The former value of $Y_p$ is consistent with the value inferred from the WMAP data 
analysis, implying the validity of the Standard Big Bang Nucleosynthesis (SBBN)
model, while the latter value of $Y_p$ is higher by more than 2$\sigma$, 
implying that some deviations from the SBBN could be present.
Treating the different He {\sc i} line emissivities as systematics
affecting the $Y_{wm}$ and $Y_p$ determinations, we obtain
$Y_{wm}$ = 0.2500 $\pm$ 0.0012 (stat.) $\pm$ 0.0015 (syst.),
and $Y_p$ = 0.2493 $\pm$ 0.0011 (stat.) $\pm$ 0.0020 (syst.). 
Both $Y_{wm}$ and $Y_p$ values within the errors agree with the 
$Y_p$ value from the WMAP data analysis.

3. The abundance ratios of the $\alpha$-elements to oxygen in both galaxies
follow the trends found in previous studies of low-metallicity emission-line 
galaxies. In particular, new data confirm findings by \citet{Iz06} that
Ne/O increases with increasing oxygen abundance, implying a higher
depletion of oxygen in higher-metallicity galaxies. We find that S/O
in both galaxies is close to the average value obtained for other 
low-metallicity emission-line galaxies. On the other hand, Cl/O in
SBS 0335--052E is lower than the average value for other low-metallicity
emission-line galaxies, while Ar/O is higher than the average value.
These differences could be due to the metallicity-dependent yields of
$\alpha$-elements synthesized by massive stars. However, it is also
possible that differences are caused by uncertainties in the atomic
data for some elements (e.g., uncertainties in the dielectronic recombination
rates), which infer incorrect abundances of sulfur, chlorine, and 
argon.

4. The measurement of Fe/O in SBS 0335--052E is among the highest values
found in emission-line
galaxies and follows the general trend of decreasing Fe/O with increasing
oxygen abundance implying iron depletion onto dust \citep{Iz06}. 
Despite the detection of dust in SBS 0335--052E by \citet{T99} 
and \citet{Houck04}, the high gas-phase Fe/O abundance ratio
suggests that iron is hardly depleted in the ionized medium of this galaxy.

5. We find that N/O in SBS 0335--052E and SBS 0335--052W is slightly
above the plateau value found by \citet{IT99} and \citet{Iz06} for
extremely low-metallicity emission-line galaxies.
There is a tendency for N/O increasing with decreasing oxygen abundance in
extremely low-metallicity galaxies with 12 + log O/H $<$ 7.6, 
probably because of an enhanced production of primary nitrogen by 
lower-metallicity rapidly rotating stars \citep{MM02}.

\begin{acknowledgements}
N. G. G. and Y. I. I. thank the hospitality of the Max Planck Institute for 
Radioastronomy (Bonn), and acknowledge support through DFG grant 
No. FR 325/57-1. This research was partially funded by project grant
AYA2007-67965-C03-02 of the Spanish Ministerio de Ciencia e Innovacion.
\end{acknowledgements}

\Online

\renewcommand{\baselinestretch}{1.0}

\begin{table*}
\caption{Extinction-corrected emission line fluxes (low-resolution FORS observations) \label{tab1}}
\begin{tabular}{lrr} \hline
 &  \multicolumn{2}{c}{0335--052E} \\ \cline{2-3} 
Line                            & No. 1+2         &  No. 7          \\ \hline \hline
3727 [O {\sc ii}]               &23.49$\pm$0.55 &24.82$\pm$2.84 \\
3750 H12                        &3.90$\pm$0.42  &     ...       \\
3771 H11                        &4.60$\pm$0.43  &     ...       \\
3797 H10                        &5.89$\pm$0.37  &     ...       \\
3820 He {\sc i}                 &1.04$\pm$0.21  &     ...       \\
3835 H9                         &8.10$\pm$0.35  &9.34$\pm$2.15  \\
3869 [Ne {\sc iii}]             &24.46$\pm$0.43 &17.48$\pm$1.39 \\
3889 He {\sc i} + H8            &16.77$\pm$0.40 &22.70$\pm$1.85 \\
3968 [Ne {\sc iii}] + H7        &24.22$\pm$0.47 &22.06$\pm$1.65 \\
4026 He {\sc i}                 &1.64$\pm$0.11  &      ...      \\
4069 [S  {\sc ii}]              &0.41$\pm$0.08  &      ...      \\
4076 [S  {\sc ii}]              &     ...       &      ...      \\
4101 H$\delta$                  &26.30$\pm$0.47 &27.73$\pm$1.63 \\
4227 [Fe {\sc iv}]              &1.21$\pm$0.45  &      ...      \\
4340 H$\gamma$                  &47.38$\pm$0.74 &47.06$\pm$1.60 \\
4363 [O {\sc iii}]              &11.22$\pm$0.19 &6.88$\pm$0.58  \\
4388 He {\sc i}                 &0.40$\pm$0.08  &      ...      \\
4471 He {\sc i}                 &3.90$\pm$0.10  &3.76$\pm$0.52  \\
4658 [Fe {\sc iii}]             &0.37$\pm$0.05  &      ...      \\
4686 He {\sc ii}                &1.16$\pm$0.06  &      ...      \\
4711 [Ar {\sc iv}] + He {\sc i} &1.68$\pm$0.07  &      ...      \\
4740 [Ar {\sc iv}]              & 0.86$\pm$0.06 &      ...      \\
4861 H$\beta$                   &100.00$\pm$1.46&100.00$\pm$2.00\\
4921 He {\sc i}                 &0.96$\pm$0.05  &      ...      \\
4959 [O {\sc iii}]              &109.47$\pm$1.58&68.06$\pm$1.25 \\
4988 [Fe {\sc iii}]             &0.47$\pm$0.05  &      ...      \\
5007 [O {\sc iii}]              &315.18$\pm$4.54&198.83$\pm$3.33\\
5199 [N {\sc i}]                &0.26$\pm$0.05  &      ...      \\
5518 [Cl {\sc iii}]             &0.14$\pm$0.03  &      ...      \\
5538 [Cl {\sc iii}]             &0.08$\pm$0.03  &      ...      \\
5876 He {\sc i}                 &10.82$\pm$0.17 &9.22$\pm$0.39  \\
6300 [O {\sc i}]                &0.75$\pm$0.04  &      ...      \\
6312 [S {\sc iii}]              &0.61$\pm$0.04  &      ...      \\
6364 [O {\sc i}]                &0.24$\pm$0.03  &      ...      \\
6563 H$\alpha$                  &273.80$\pm$4.28&273.83$\pm$4.90\\
6583 [N {\sc ii}]               &0.90$\pm$0.04  &1.01$\pm$0.31  \\
6678 He  {\sc i}                &2.57$\pm$0.06  &2.74$\pm$0.29  \\
6716 [S {\sc ii}]               &1.74$\pm$0.05  &2.59$\pm$0.32  \\
6731 [S {\sc ii}]               &1.65$\pm$0.05  &2.48$\pm$0.35  \\
7065 He  {\sc i}                &4.29$\pm$0.09  &1.89$\pm$0.29  \\
7136 [Ar {\sc iii}]             &1.52$\pm$0.05  &1.07$\pm$0.24  \\
7281 He  {\sc i}                &0.62$\pm$0.04  &      ...      \\
7320 [O {\sc ii}]               &0.58$\pm$0.04  &      ...      \\
7330 [O {\sc ii}]               &0.43$\pm$0.04  &      ...      \\
$C$(H$\beta$)                   &    0.180      &     0.140     \\
EW(H$\beta$))$^a$               &     351       &      298      \\
$F$(H$\beta$)$^b$               &     222.2     &      16.0     \\
EW(abs)                         &     2.5       &      4.2      \\
\hline
\end{tabular}

$^a$ in $\AA$. \\
$^b$ in units 10$^{-16}$ erg s$^{-1}$ cm$^{-2}$.
\end{table*}

\begin{table*}
\caption{Extinction-corrected emission line fluxes (high-resolution FORS observations) \label{tab2}}
\begin{tabular}{lrrrrrrr} \hline
 &  \multicolumn{2}{c}{0335--052E} &&  \multicolumn{4}{c}{0335--052W} \\ \cline{2-3} \cline{5-8} 
Line                            & No. 1+2         &  No. 7          &&         No. 1   &         No. 2   &        No. 3     &         No. 4     \\ \hline\hline
3727 [O {\sc ii}]               &21.81$\pm$0.34 &24.98$\pm$0.53 &&75.20$\pm$2.55 &30.15$\pm$2.20 &79.16$\pm$7.95  &41.95$\pm$4.46 \\
3750 H12                        &3.08$\pm$0.07  &5.84$\pm$0.49  &&      ...      &      ...      &      ...       &     ...       \\
3771 H11                        &3.92$\pm$0.08  &6.32$\pm$0.45  &&      ...      &      ...      &      ...       &     ...       \\
3797 H10                        &5.39$\pm$0.10  &7.28$\pm$0.43  &&5.65$\pm$0.11  &9.14$\pm$1.96  &      ...       &     ...       \\
3820 He {\sc i}                 &1.06$\pm$0.04  &0.91$\pm$0.19  &&      ...      &      ...      &      ...       &     ...       \\
3835 H9                         &7.30$\pm$0.13  &9.16$\pm$0.39  &&7.02$\pm$0.13  &9.96$\pm$1.81  &      ...       &     ...       \\
3869 [Ne {\sc iii}]             &26.98$\pm$0.41 &15.86$\pm$0.35 &&11.31$\pm$0.98 &12.43$\pm$1.07 &      ...       &     ...       \\
3889 He {\sc i} + H8            &16.96$\pm$0.26 &20.75$\pm$0.48 &&19.92$\pm$1.02 &24.69$\pm$1.61 &      ...       &24.83$\pm$3.90 \\
3968 [Ne {\sc iii}] + H7        &25.56$\pm$0.39 &23.30$\pm$0.50 &&13.52$\pm$0.96 &23.38$\pm$1.51 &      ...       &25.28$\pm$4.25 \\
4026 He {\sc i}                 &1.79$\pm$0.04  &1.94$\pm$0.19  &&      ...      &      ...      &      ...       &     ...       \\
4069 [S  {\sc ii}]              &0.39$\pm$0.03  &      ...      &&      ...      &      ...      &      ...       &     ...       \\
4076 [S  {\sc ii}]              &     ...       &      ...      &&      ...      &      ...      &      ...       &     ...       \\
4101 H$\delta$                  &27.11$\pm$0.40 &27.10$\pm$0.52 &&26.61$\pm$1.09 &27.17$\pm$1.51 &32.76$\pm$6.79  &24.80$\pm$3.80 \\
4227 [Fe {\sc iv}]              &0.25$\pm$0.04  &      ...      &&      ...      &      ...      &      ...       &     ...        \\ 
4340 H$\gamma$                  &47.61$\pm$0.69 &46.39$\pm$0.76 &&45.94$\pm$1.28 &46.85$\pm$1.50 &44.10$\pm$4.96  &43.27$\pm$2.76 \\
4363 [O {\sc iii}]              &11.33$\pm$0.17 &6.15$\pm$0.18  &&3.33$\pm$0.58  &5.09$\pm$0.72  &      ...       &2.51$\pm$0.80  \\
4388 He {\sc i}                 &0.40$\pm$0.02  &      ...      &&      ...      &      ...      &      ...       &     ...        \\ 
4471 He {\sc i}                 &3.77$\pm$0.06  &3.54$\pm$0.13  &&3.55$\pm$0.32  &4.52$\pm$0.71  &      ...       &     ...        \\
4658 [Fe {\sc iii}]             &0.32$\pm$0.02  &0.40$\pm$0.09  &&      ...      &      ...      &      ...       &     ...       \\
4686 He {\sc ii}                &1.33$\pm$0.03  &1.33$\pm$0.14  &&      ...      &      ...      &      ...       &     ...        \\
4711 [Ar {\sc iv}] + He {\sc i} &1.72$\pm$0.03  &0.80$\pm$0.09  &&      ...      &      ...      &      ...       &     ...        \\
4740 [Ar {\sc iv}]              & 0.92$\pm$0.02 &0.44$\pm$0.12  &&      ...      &      ...      &      ...       &     ...        \\
4861 H$\beta$                   &100.00$\pm$1.42&100.00$\pm$1.51&&100.00$\pm$2.01&100.00$\pm$2.34&100.00$\pm$5.42 &100.00$\pm$3.40 \\
4921 He {\sc i}                 &0.91$\pm$0.02  &0.87$\pm$0.10  &&      ...      &      ...      &      ...       &     ...        \\
4959 [O {\sc iii}]              &101.50$\pm$1.44&64.87$\pm$0.99 &&43.23$\pm$1.02 &49.24$\pm$1.33 &22.91$\pm$2.56  &23.50$\pm$1.43 \\
4988 [Fe {\sc iii}]             &0.38$\pm$0.01  &0.60$\pm$0.08  &&      ...      &      ...      &      ...       &     ...        \\ 
5007 [O {\sc iii}]              &303.19$\pm$4.31&194.48$\pm$2.89&&130.20$\pm$2.52&147.95$\pm$3.21&71.23$\pm$3.74  &75.37$\pm$2.42 \\
5016 He {\sc i}                 &2.09$\pm$0.03  &2.06$\pm$0.10  &&2.99$\pm$0.41  &2.50$\pm$0.58  &      ...       &     ...       \\
5199 [N {\sc i}]                &0.28$\pm$0.01  &      ...      &&      ...      &      ...      &      ...       &     ...       \\
5518 [Cl {\sc iii}]             &0.09$\pm$0.02  &      ...      &&      ...      &      ...      &      ...       &     ...        \\
5538 [Cl {\sc iii}]             &0.07$\pm$0.02  &      ...      &&      ...      &      ...      &      ...       &     ...        \\
5755 [N {\sc ii}]               &0.06$\pm$0.01  &      ...      &&      ...      &      ...      &      ...       &     ...        \\
5876 He {\sc i}                 &10.48$\pm$0.16 &9.09$\pm$0.22  &&9.02$\pm$0.75  &10.34$\pm$0.87 &      ...       &     ...        \\
6300 [O {\sc i}]                &0.69$\pm$0.02  &0.75$\pm$0.12  &&      ...      &      ...      &      ...       &     ...        \\
6312 [S {\sc iii}]              &0.59$\pm$0.02  &0.51$\pm$0.12  &&      ...      &      ...      &      ...       &     ...        \\
6364 [O {\sc i}]                &0.22$\pm$0.01  &      ...      &&      ...      &      ...      &      ...       &     ...        \\
6548 [N {\sc ii}]               &0.28$\pm$0.02  &      ...      &&      ...      &      ...      &      ...       &     ...        \\
6563 H$\alpha$                  &272.75$\pm$4.20&274.56$\pm$4.43&&279.19$\pm$5.65&274.11$\pm$6.27&289.06$\pm$11.70&273.47$\pm$8.17 \\
6583 [N {\sc ii}]               &0.97$\pm$0.02  &0.85$\pm$0.13  &&3.40$\pm$0.49  &      ...      &      ...       &     ...        \\
6678 He  {\sc i}                &2.63$\pm$0.05  &2.55$\pm$0.15  &&2.54$\pm$0.44  &      ...      &      ...       &     ...        \\
6716 [S {\sc ii}]               &1.89$\pm$0.04  &2.58$\pm$0.13  &&6.89$\pm$0.59  &      ...      &      ...       &     ...        \\
6731 [S {\sc ii}]               &1.62$\pm$0.03  &1.95$\pm$0.14  &&4.08$\pm$0.48  &      ...      &      ...       &     ...        \\
7065 He  {\sc i}                &4.46$\pm$0.08  &2.15$\pm$0.13  &&2.20$\pm$0.41  &      ...      &      ...       &     ...        \\
7136 [Ar {\sc iii}]             &1.57$\pm$0.03  &1.47$\pm$0.14  &&      ...      &      ...      &      ...       &     ...        \\
7281 He  {\sc i}                &0.60$\pm$0.02  &0.75$\pm$0.14  &&      ...      &      ...      &      ...       &     ...        \\
7320 [O {\sc ii}]               &0.57$\pm$0.02  &0.77$\pm$0.18  &&      ...      &      ...      &      ...       &     ...        \\
7330 [O {\sc ii}]               &0.41$\pm$0.02  &0.34$\pm$0.11  &&      ...      &      ...      &      ...       &     ...        \\
$C$(H$\beta$)                   &    0.135      &     0.090     &&     0.485     &    0.180      &     0.115      &    0.040       \\
EW(H$\beta$))$^a$                    &     337       &      305      &&      93       &     161       &      53        &     103        \\
$F$(H$\beta$)$^b$               &     220.3     &      16.2     &&      1.4      &     1.2       &      0.5       &     0.8        \\
EW(abs)                         &     0.0       &      4.6      &&      0.0      &     4.0       &      2.0       &     5.8        \\
\hline
\end{tabular}

$^a$ in $\AA$. \\
$^b$ in units 10$^{-16}$ erg s$^{-1}$ cm$^{-2}$.
\end{table*}

\begin{table*}
\caption{Extinction-corrected weak emission line fluxes in 
SBS 0335--052E No. 1+2 (high-resolution FORS observations)\label{tab3}}
\begin{tabular}{lrclrclrclr} \hline
Line & Flux$^a$ & & Line & Flux$^a$ & & Line & Flux$^a$ & & Line & Flux$^a$ \\ \hline
3530 He {\sc i}       & 0.10$\pm$0.10&& 4169 He {\sc i}+O {\sc ii}& 0.06$\pm$0.02&& 4650 O {\sc ii}                  & 0.02$\pm$0.01&& 5261 [Fe {\sc ii}]   & 0.06$\pm$0.01 \\
3554 He {\sc i}       & 0.22$\pm$0.06&& 4201 He {\sc ii}          & 0.02$\pm$0.01&& 4702 [Fe {\sc iii}]              & 0.11$\pm$0.01&& 5271 [Fe {\sc iii}]  & 0.16$\pm$0.01 \\
3587 He {\sc i}       & 0.31$\pm$0.05&& 4245 [Fe {\sc ii}]        & 0.03$\pm$0.01&& 4723 [Ne {\sc iv}]               & 0.02$\pm$0.01&& 5297 O {\sc i}       & 0.04$\pm$0.01 \\
3614 He {\sc i}       & 0.22$\pm$0.04&& 4267 C {\sc ii}           & 0.02$\pm$0.01&& 4734 [Fe {\sc iii}]              & 0.03$\pm$0.01&& 5411 He {\sc ii}     & 0.10$\pm$0.01 \\
3635 He {\sc i}       & 0.45$\pm$0.06&& 4287 [Fe {\sc ii}]        & 0.15$\pm$0.02&& 4755 [Fe {\sc iii}]              & 0.06$\pm$0.01&& 5513 O {\sc i}       & 0.03$\pm$0.01 \\
3683 H20              & 0.20$\pm$0.02&& 4320 O {\sc ii}           & 0.03$\pm$0.01&& 4769 [Fe {\sc iii}]              & 0.03$\pm$0.01&& 5577 O {\sc i}       & 0.01$\pm$0.01 \\
3687 H19              & 0.38$\pm$0.03&& 4348 O {\sc ii}           & 0.04$\pm$0.01&& 4788 N {\sc ii}                  & 0.05$\pm$0.01&& 5958 Si{\sc ii}      & 0.09$\pm$0.01 \\
3692 H18              & 0.66$\pm$0.03&& 4414 [Fe {\sc ii}]        & 0.15$\pm$0.02&& 4815 [Fe {\sc ii}]+S{\sc ii}     & 0.07$\pm$0.01&& 5979 Si{\sc ii}      & 0.07$\pm$0.01 \\
3697 H17              & 1.00$\pm$0.04&& 4438 He {\sc i}           & 0.09$\pm$0.02&& 4881 [Fe {\sc iii}]              & 0.12$\pm$0.02&& 6046 O {\sc i}       & 0.06$\pm$0.01 \\
3704 H16              & 1.82$\pm$0.04&& 4452 [Fe {\sc ii}]        & 0.06$\pm$0.02&& 4890 [Fe {\sc ii}]               & 0.09$\pm$0.02&& 6347 Si{\sc ii}      & 0.11$\pm$0.01 \\
3712 H15              & 1.57$\pm$0.04&& 4510 N {\sc iii}          & 0.03$\pm$0.01&& 4931 [O {\sc iii}]               & 0.05$\pm$0.01&& 6371 Si{\sc ii}      & 0.09$\pm$0.01 \\
3722 H14+[S {\sc iii}]& 2.31$\pm$0.04&& 4514 N {\sc iii}          & 0.03$\pm$0.01&& 5041 Si {\sc ii}                 & 0.20$\pm$0.01&& 7002 O {\sc i}       & 0.08$\pm$0.02 \\
3734 H13              & 2.24$\pm$0.04&& 4536 N {\sc iii}          & 0.01$\pm$0.01&& 5047 He {\sc i}                  & 0.20$\pm$0.01&& 7170 [Ar {\sc iv}]   & 0.06$\pm$0.02 \\
3856 Si {\sc ii}      & 0.07$\pm$0.02&& 4549 N {\sc iii}          & 0.04$\pm$0.02&& 5056 Si {\sc ii}                 & 0.08$\pm$0.01&& 7237 [Ar {\sc iv}]   & 0.04$\pm$0.01 \\
3927 He {\sc i}       & 0.09$\pm$0.02&& 4563 [Mg {\sc i}]         & 0.03$\pm$0.01&& 5112 [Fe {\sc ii}]               & 0.04$\pm$0.01&& 7263 [Ar {\sc iv}]   & 0.05$\pm$0.01 \\
4009 He {\sc i}       & 0.18$\pm$0.02&& 4571 Mg {\sc i}           & 0.05$\pm$0.02&& 5146 [Fe {\sc vi}]               & 0.03$\pm$0.01 \\
4121 He {\sc i}       & 0.26$\pm$0.02&& 4624 N {\sc ii}           & 0.02$\pm$0.01&& 5159 [Fe {\sc ii}]+[Fe {\sc vii}]& 0.08$\pm$0.01 \\
4143 He {\sc i}       & 0.27$\pm$0.02&& 4634 N {\sc iii}          & 0.01$\pm$0.01&& 5176 [Fe {\sc vi}]               & 0.04$\pm$0.01 \\
4154 O {\sc ii}       & 0.06$\pm$0.02&& 4642 N {\sc iii}          & 0.03$\pm$0.01&& 5191 [Ar {\sc iii}]              & 0.04$\pm$0.01 \\ \hline \\
\end{tabular}

$^a$ $I$(H$\beta$)=100. \\
\end{table*}

\begin{table*}
\caption{Extinction-corrected emission line fluxes (UVES observations) \label{tab4}}
\begin{tabular}{lrrrrrrrr} \hline
 &  \multicolumn{6}{c}{0335--052E} & & \\ \cline{2-7}  
Line                            & No. 1+2$^a$     &  No. 1+2$^b$    &   No. 4+5$^b$   &   No. 4+5$^c$   &No. (1+2)+(4+5)$^b$& No. 7$^c$  && 0335--052W$^b$  \\ \hline\hline

3346 [Ne {\sc v}]               &     ...       &     ...       &     ...       &      ...      &0.24$\pm$0.00   &     ...       &&      ...     \\
3425 [Ne {\sc v}]               &0.97$\pm$0.17  &     ...       &1.74$\pm$0.03  &      ...      &0.57$\pm$0.01   &     ...       &&      ...     \\
3727 [O {\sc ii}]               &24.84$\pm$0.39 &27.22$\pm$0.43 &24.03$\pm$0.44 &25.26$\pm$0.54 &26.19$\pm$0.41  &18.79$\pm$0.30  && 52.06$\pm$0.89\\
3750 H12                        &4.14$\pm$0.08  &4.17$\pm$0.08  &2.75$\pm$0.15  &4.57$\pm$0.31  &3.47$\pm$0.09   &8.32$\pm$1.27  && 5.60$\pm$0.27\\
3771 H11                        &5.08$\pm$0.09  &4.81$\pm$0.09  &3.64$\pm$0.17  &5.01$\pm$0.28  &4.23$\pm$0.09   &5.73$\pm$3.23  && 6.22$\pm$0.26\\
3797 H10                        &6.41$\pm$0.11  &6.21$\pm$0.11  &4.79$\pm$0.18  &6.68$\pm$0.31  &5.51$\pm$0.11   &8.82$\pm$1.47  && 7.53$\pm$0.27\\
3820 He {\sc i}                 &0.93$\pm$0.03  &0.92$\pm$0.03  &      ...      &      ...      &0.83$\pm$0.05   &     ...       &&       ...    \\
3835 H9                         &9.30$\pm$0.15  &8.28$\pm$0.14  &6.51$\pm$0.19  &9.30$\pm$0.32  &7.61$\pm$0.13   &10.96$\pm$1.21 && 9.20$\pm$0.26\\
3869 [Ne {\sc iii}]             &25.02$\pm$0.38 &23.41$\pm$0.36 &21.75$\pm$0.37 &25.21$\pm$0.46 &23.37$\pm$0.36  &12.24$\pm$0.59 &&9.19$\pm$0.20\\
3889 He {\sc i} + H8            &18.24$\pm$0.28 &17.04$\pm$0.27 &16.63$\pm$0.34 &17.80$\pm$0.42 &16.46$\pm$0.27  &16.36$\pm$0.94 &&20.05$\pm$0.42 \\
3968 [Ne {\sc iii}] + H7        &     ...       &     ...       &      ...      &      ...      &      ...       &7.40$\pm$1.08  &&     ...      \\
4026 He {\sc i}                 &1.70$\pm$0.04  &1.63$\pm$0.04  &1.34$\pm$0.11  &1.64$\pm$0.17  &1.67$\pm$0.06   &     ...       && 1.31$\pm$0.12\\
4069 [S  {\sc ii}]              &0.36$\pm$0.02  &0.31$\pm$0.02  &      ...      &      ...      &0.30$\pm$0.03   &     ...       &&     ...      \\
4076 [S  {\sc ii}]              &0.13$\pm$0.01  &      ...      &      ...      &      ...      &      ...       &     ...       &&     ...      \\
4101 H$\delta$                  &27.92$\pm$0.42 &26.22$\pm$0.39 &23.67$\pm$0.40 &28.84$\pm$0.53 &25.64$\pm$0.39  &24.88$\pm$0.92 && 26.59$\pm$0.46\\
4227 [Fe {\sc v}]               &0.81$\pm$0.06  &1.34$\pm$0.11  &      ...      &      ...      &0.76$\pm$0.12   &     ...       &&     ...       \\ 
4340 H$\gamma$                  &48.02$\pm$0.70 &47.97$\pm$0.70 &48.77$\pm$0.73 &49.42$\pm$0.77 &48.31$\pm$0.70  &46.87$\pm$1.00 && 47.41$\pm$0.73\\
4363 [O {\sc iii}]              &10.84$\pm$0.16 &10.92$\pm$0.17 &11.22$\pm$0.23 &12.49$\pm$0.28 &11.11$\pm$0.18  &7.55$\pm$0.65  && 3.92$\pm$0.15\\
4388 He {\sc i}                 &0.41$\pm$0.02  &0.51$\pm$0.02  &      ...      &      ...      &0.44$\pm$0.02   &     ...       &&      ...     \\ 
4471 He {\sc i}                 &3.58$\pm$0.06  &     ...       &      ...      &3.43$\pm$0.12  &      ...       &     ...       &&      ...     \\
4658 [Fe {\sc iii}]             &0.34$\pm$0.02  &     ...       &      ...      &      ...      &      ...       &     ...       &&      ...     \\
4686 He {\sc ii}                &2.78$\pm$0.05  &    ...        &      ...      &4.36$\pm$0.17  &      ...       &     ...       &&      ...     \\
4711 [Ar {\sc iv}] + He {\sc i} &      ...      &    ...        &      ...      &      ...      &      ...       &     ...       &&      ...      \\
4740 [Ar {\sc iv}]              & 0.93$\pm$0.03 &0.90$\pm$0.01  &0.84$\pm$0.03  &1.12$\pm$0.05  &0.85$\pm$0.02   &     ...       &&      ...      \\
4861 H$\beta$                   &100.00$\pm$1.43&100.00$\pm$1.43&100.00$\pm$1.43&100.00$\pm$1.45&100.00$\pm$1.42 &100.00$\pm$1.56&& 100.00$\pm$1.46\\
4921 He {\sc i}                 &0.81$\pm$0.03  &0.93$\pm$0.02  &0.71$\pm$0.02  &0.94$\pm$0.06  &0.85$\pm$0.02   &     ...       && 0.97$\pm$0.04\\
4959 [O {\sc iii}]              &101.66$\pm$1.45&103.51$\pm$1.48&110.60$\pm$1.58&111.20$\pm$1.61&107.84$\pm$1.53 &82.38$\pm$1.32 &&45.84$\pm$0.68\\
4988 [Fe {\sc iii}]             &      ...      &      ...      &      ...      &      ...      &      ...       &     ...       &&      ...     \\ 
5007 [O {\sc iii}]              &311.02$\pm$4.44&309.37$\pm$4.41&337.19$\pm$4.81&332.42$\pm$4.79&326.75$\pm$4.65 &247.58$\pm$3.81&&137.25$\pm$2.00\\
5016 He {\sc i}                 &1.86$\pm$0.04  &1.77$\pm$0.03  &1.73$\pm$0.03  &1.71$\pm$0.06  &1.70$\pm$0.03   &1.79$\pm$0.13  && 1.96$\pm$0.04\\
5199 [N {\sc i}]                &     ...       &      ...      &      ...      &      ...      &      ...       &     ...       &&      ...     \\
5518 [Cl {\sc iii}]             &0.13$\pm$0.03  &0.11$\pm$0.01  &      ...      &      ...      &0.12$\pm$0.01   &     ...       &&      ...     \\
5538 [Cl {\sc iii}]             &0.05$\pm$0.02  &0.06$\pm$0.00  &      ...      &      ...      &0.07$\pm$0.01   &     ...       &&      ...     \\
5755 [N {\sc ii}]               &      ...      &      ...      &      ...      &      ...      &      ...       &     ...       &&      ...     \\
5876 He {\sc i}                 &9.22$\pm$0.15  &11.70$\pm$0.18 &8.54$\pm$0.13  &9.26$\pm$0.17  &10.74$\pm$0.16  &9.78$\pm$0.31  && 8.88$\pm$0.15\\
6300 [O {\sc i}]                &0.67$\pm$0.03  &0.86$\pm$0.02  &0.60$\pm$0.02  &0.75$\pm$0.04  &0.80$\pm$0.02   &0.96$\pm$0.11  && 1.23$\pm$0.04\\
6312 [S {\sc iii}]              &0.55$\pm$0.03  &0.67$\pm$0.02  &0.56$\pm$0.02  &0.64$\pm$0.06  &0.65$\pm$0.02   &     ...       && 0.74$\pm$0.05\\
6364 [O {\sc i}]                &0.23$\pm$0.02  &0.29$\pm$0.01  &0.22$\pm$0.02  &0.21$\pm$0.03  &0.27$\pm$0.01   &     ...       && 0.36$\pm$0.03\\
6548 [N {\sc ii}]               &0.26$\pm$0.02  &0.42$\pm$0.01  &0.27$\pm$0.01  &0.28$\pm$0.02  &0.36$\pm$0.01   &     ...       && 0.60$\pm$0.02\\
6563 H$\alpha$                  &273.64$\pm$4.24&272.68$\pm$4.22&274.28$\pm$4.24&272.73$\pm$4.25&273.10$\pm$4.21 &274.73$\pm$4.52&& 275.55$\pm$4.34\\
6583 [N {\sc ii}]               &0.94$\pm$0.04  &1.19$\pm$0.02  &0.79$\pm$0.03  &0.86$\pm$0.05  &1.08$\pm$0.03   &0.77$\pm$0.14  && 1.91$\pm$0.06\\
6678 He  {\sc i}                &2.62$\pm$0.05  &3.04$\pm$0.05  &2.49$\pm$0.05  &2.56$\pm$0.05  &2.97$\pm$0.05   &2.58$\pm$0.09  && 2.41$\pm$0.06\\
6716 [S {\sc ii}]               &1.89$\pm$0.05  &2.01$\pm$0.03  &1.82$\pm$0.04  &1.62$\pm$0.04  &1.98$\pm$0.03   &2.10$\pm$0.08  && 4.97$\pm$0.10\\
6731 [S {\sc ii}]               &1.34$\pm$0.05  &    ...        &     ...       &1.32$\pm$0.04  &      ...       &1.41$\pm$0.07  &&      ...     \\
7065 He  {\sc i}                &4.17$\pm$0.07  &    ...        &     ...       &2.75$\pm$0.06  &      ...       &2.02$\pm$0.10  &&      ...    \\
7136 [Ar {\sc iii}]             &1.88$\pm$0.03  &    ...        &      ...      &1.45$\pm$0.04  &      ...       &1.23$\pm$0.12  &&      ...    \\
7281 He  {\sc i}                &0.69$\pm$0.02  &    ...        &      ...      &0.57$\pm$0.03  &      ...       &     ...       &&      ...    \\
7320 [O {\sc ii}]               &0.64$\pm$0.02  &    ...        &      ...      &0.38$\pm$0.03  &      ...       &     ...       &&      ...    \\
7330 [O {\sc ii}]               &0.54$\pm$0.01  &    ...        &      ...      &0.38$\pm$0.03  &      ...       &     ...       &&      ...    \\
9069 [S {\sc iii}]              &4.63$\pm$0.14  &    ...        &      ...      &2.80$\pm$0.18  &      ...       &2.34$\pm$0.06  &&      ...    \\
$C$(H$\beta$)                   &    0.140      &     0.010     &     0.140     &    0.140      &     0.010      &    0.050      &&      0.055  \\
EW(H$\beta$))$^d$               &     201       &      332      &      104      &     108       &      182       &     280       &&      122    \\
$F$(H$\beta$)$^e$               &     747.7     &      368.4    &      153.3    &     256.7     &      499.6     &     39.2       &&      14.5    \\
EW(abs)                         &     0.8       &      2.2      &      0.3      &     0.2       &      0.7       &     5.3       &&      2.5    \\
\hline
\end{tabular}

$^a$ ESO program 71.B-0055(A) \\
$^b$ ESO program 70.B-0717(A) \\
$^c$ ESO program 68.B-0310(A)  \\
$^d$ in $\AA$. \\
$^e$ in units 10$^{-16}$ erg s$^{-1}$ cm$^{-2}$.

\end{table*}

\begin{table*}[t]
\caption{Ionic and total element abundances (low-resolution FORS observations) \label{tab5}}
\begin{tabular}{lcc} \hline
 &  \multicolumn{2}{c}{0335--052E} \\ \cline{2-3} 
Property                            & No. 1+2         &  No. 7       \\ \hline\hline

$T_e$(O {\sc iii}), K           &20549$\pm$264  &20243$\pm$1058   \\
$T_e$(O {\sc ii}), K            &16356$\pm$185  &16320$\pm$753   \\
$T_e$(S {\sc iii}), K           &18755$\pm$219  &18502$\pm$878   \\
$N_e$([S {\sc ii}]), cm$^{-3}$  & 541$\pm$97    & 564$\pm$525    \\ \\
                                                            
O$^+$/H$^+$, ($\times$10$^5$)     &0.174$\pm$0.006&0.192$\pm$0.030 \\
O$^{2+}$/H$^+$, ($\times$10$^5$)  &1.722$\pm$0.053&1.116$\pm$0.132 \\
O$^{3+}$/H$^+$, ($\times$10$^6$)  &0.220$\pm$0.013&       ...      \\
O/H, ($\times$10$^5$)             &1.918$\pm$0.053&1.308$\pm$0.135 \\
12+log O/H                        &7.28$\pm$0.01  &7.12$\pm$0.04   \\ \\
Ne$^{2+}$/H$^+$, ($\times$10$^6$) &2.912$\pm$0.092&2.153$\pm$0.289 \\
$ICF$(Ne)                         &      1.03     &      1.05      \\
Ne/H, ($\times$10$^6$)            &3.011$\pm$0.102&2.255$\pm$0.339 \\
log Ne/O                          &--0.80$\pm$0.02&--0.76$\pm$0.08 \\ \\
N$^{+}$/H$^+$, ($\times$10$^7$)   &0.563$\pm$0.021&0.633$\pm$0.158 \\
$ICF$(N)                          &      10.21    &      6.52      \\
N/H, ($\times$10$^7$)             &5.818$\pm$0.234&4.126$\pm$1.077 \\
log N/O                           &--1.52$\pm$0.02&--1.50$\pm$0.12 \\ \\
S$^{+}$/H$^+$, ($\times$10$^7$)   &0.298$\pm$0.008&0.446$\pm$0.056 \\
S$^{2+}$/H$^+$, ($\times$10$^7$)  &1.680$\pm$0.111&       ...      \\
$ICF$(S)                          &      2.22     &       ...      \\
S/H, ($\times$10$^7$)             &4.440$\pm$0.248&       ...      \\
log S/O                           &--1.64$\pm$0.03&       ...      \\ \\
Cl$^{2+}$/H$^+$, ($\times$10$^8$) &0.348$\pm$0.075&      ...       \\
$ICF$(Cl)                         &      1.38     &      ...       \\
Cl/H, ($\times$10$^8$)            &0.480$\pm$0.103&      ...       \\
log Cl/O                          &--3.60$\pm$0.09&      ...       \\ \\
Ar$^{2+}$/H$^+$, ($\times$10$^7$) &0.426$\pm$0.014&0.308$\pm$0.069 \\
Ar$^{3+}$/H$^+$, ($\times$10$^7$) &0.413$\pm$0.028&       ...      \\
$ICF$(Ar)                         &      2.14     &      1.53      \\
Ar/H, ($\times$10$^7$)            &0.912$\pm$0.068&0.471$\pm$0.106 \\
log Ar/O                          &--2.32$\pm$0.03&--2.44$\pm$0.11 \\ \\
Fe$^{2+}$/H$^+$, ($\times$10$^7$)(4658)&0.572$\pm$0.073&   ...     \\
$ICF$(Fe)                              &      15.25    &   ...     \\
Fe/H, ($\times$10$^7$)(4658)           &8.729$\pm$1.119&   ...     \\
log Fe/O (4658)                        &--1.34$\pm$0.06&   ...     \\ \\
Fe$^{2+}$/H$^+$, ($\times$10$^7$)(4988)&0.724$\pm$0.073&   ...     \\
$ICF$(Fe)                              &      15.25    &   ...     \\
Fe/H, ($\times$10$^7$)(4988)           &11.050$\pm$1.118&   ...     \\
log Fe/O (4988)                        &--1.24$\pm$0.05&   ...     \\
\hline
\end{tabular}
\end{table*}

\begin{table*}[t]
\caption{Ionic and total element abundances (high-resolution FORS observations) \label{tab6}}
\begin{tabular}{lccccccc} \hline
 &  \multicolumn{2}{c}{0335--052E} &&  \multicolumn{4}{c}{0335--052W} \\ \cline{2-3} \cline{5-8} 
Property                            & No. 1+2         &  No. 7          &&         No. 1   &         No. 2   &        No. 3     &         No. 4     \\ \hline\hline

$T_e$(O {\sc iii}), K           &21314$\pm$257  &19281$\pm$352   &&17188$\pm$1557  &20273$\pm$1771 &20000$\pm$1048 &20059$\pm$3899   \\
$T_e$(O {\sc ii}), K            &16407$\pm$171  &16145$\pm$263   &&15452$\pm$1268  &16323$\pm$1259 &16284$\pm$1329 &16293$\pm$2804   \\   
$T_e$(S {\sc iii}), K           &19390$\pm$213  &17703$\pm$292   &&15966$\pm$1293  &18526$\pm$1470 &18300$\pm$870  &18349$\pm$3236   \\
$N_e$([S {\sc ii}]), cm$^{-3}$  & 316$\pm$54    & 100$\pm$118    &&     100        &      100       &     100      &      100        \\ 
                                                                                                                                     \\ 
O$^+$/H$^+$, ($\times$10$^5$)     &0.156$\pm$0.004&0.183$\pm$0.008 &&0.623$\pm$0.131&0.212$\pm$0.042&0.561$\pm$0.124&0.297$\pm$0.127 \\
O$^{2+}$/H$^+$, ($\times$10$^5$)$^a$  &1.526$\pm$0.044&1.119$\pm$0.052 &&1.032$\pm$0.226&0.822$\pm$0.162&0.403$\pm$0.052&0.421$\pm$0.185 \\
O$^{2+}$/H$^+$, ($\times$10$^5$)$^b$  &1.500$\pm$0.750&... &&...&...&...&... \\
O$^{3+}$/H$^+$, ($\times$10$^6$)  &0.230$\pm$0.009&0.204$\pm$0.025 &&      ...      &      ...      &      ...      &      ...       \\
O/H, ($\times$10$^5$)             &1.705$\pm$0.044&1.403$\pm$0.053 &&1.654$\pm$0.262&1.034$\pm$0.167&0.965$\pm$0.134&0.718$\pm$0.224 \\
12+log O/H                        &7.23$\pm$0.01  &7.15$\pm$0.02   &&7.22$\pm$0.07  &7.01$\pm$0.07  &6.98$\pm$0.06  &6.86$\pm$0.14   \\ 
                                                                                                                                     \\ 
Ne$^{2+}$/H$^+$, ($\times$10$^6$) &2.960$\pm$0.084&2.189$\pm$0.099 &&2.078$\pm$0.480&1.526$\pm$0.306&      ...      &      ...       \\
$ICF$(Ne)                         &      1.03     &      1.05      &&     1.14      &     1.07      &      ...      &      ...       \\
Ne/H, ($\times$10$^6$)            &3.064$\pm$0.094&2.291$\pm$0.116 &&2.375$\pm$0.077&1.629$\pm$0.384&      ...      &      ...       \\
log Ne/O                          &--0.74$\pm$0.02&--0.79$\pm$0.03 &&--0.84$\pm$0.16&--0.80$\pm$0.12&      ...      &      ...       \\
                                                                                                                                     \\ 
N$^{+}$/H$^+$, ($\times$10$^7$)   &0.600$\pm$0.015&0.543$\pm$0.067 &&2.364$\pm$0.426&      ...      &      ...      &      ...       \\
$ICF$(N)                          &      10.21    &      7.29      &&     2.68      &      ...      &      ...      &      ...       \\
N/H, ($\times$10$^7$)             &6.126$\pm$0.163&3.958$\pm$0.514 &&6.339$\pm$1.132&      ...      &      ...      &      ...       \\
log N/O                           &--1.44$\pm$0.02&--1.55$\pm$0.06 &&--1.42$\pm$0.10&      ...      &      ...      &      ...       \\
                                                                                                                                     \\ 
S$^{+}$/H$^+$, ($\times$10$^7$)   &0.299$\pm$0.006&0.387$\pm$0.019 &&     ...       &      ...      &      ...      &      ...       \\
S$^{2+}$/H$^+$, ($\times$10$^7$)  &1.486$\pm$0.056&1.621$\pm$0.394 &&     ...       &      ...      &      ...      &      ...       \\
$ICF$(S)                          &      2.20     &      1.72      &&     ...       &      ...      &      ...      &      ...       \\
S/H, ($\times$10$^7$)             &3.937$\pm$0.124&3.464$\pm$0.681 &&     ...       &      ...      &      ...      &      ...       \\
log S/O                           &--1.64$\pm$0.02&--1.61$\pm$0.08 &&     ...       &      ...      &      ...      &      ...       \\
                                                                                                                                     \\ 
Cl$^{2+}$/H$^+$, ($\times$10$^8$) &0.249$\pm$0.035&      ...       &&     ...       &      ...      &      ...      &      ...       \\
$ICF$(Cl)                         &      1.38     &      ...       &&     ...       &      ...      &      ...      &      ...       \\
Cl/H, ($\times$10$^8$)            &0.343$\pm$0.048&      ...       &&     ...       &      ...      &      ...      &      ...       \\
log Cl/O                          &--3.70$\pm$0.06&      ...       &&     ...       &      ...      &      ...      &      ...       \\ \\
Ar$^{2+}$/H$^+$, ($\times$10$^7$) &0.417$\pm$0.009&0.448$\pm$0.043 &&     ...       &      ...      &      ...      &      ...      \\
Ar$^{3+}$/H$^+$, ($\times$10$^7$) &0.412$\pm$0.012&0.244$\pm$0.069 &&     ...       &      ...      &      ...      &      ...       \\
$ICF$(Ar)                         &      2.12     &      1.65      &&     ...       &      ...      &      ...      &      ...       \\
Ar/H, ($\times$10$^7$)            &0.885$\pm$0.033&0.739$\pm$0.134 &&     ...       &      ...      &      ...      &      ...       \\
log Ar/O                          &--2.28$\pm$0.02&--2.28$\pm$0.08 &&     ...       &      ...      &      ...      &      ...       \\
                                                                                                                                     \\ 
Fe$^{2+}$/H$^+$, ($\times$10$^7$)(4658)&0.498$\pm$0.029&0.644$\pm$0.150 &&     ...       &      ...      &      ...      &      ...       \\
$ICF$(Fe)                              &      15.08    &      10.53     &&     ...       &      ...      &      ...      &      ...       \\
Fe/H, ($\times$10$^7$)(4658)           &7.506$\pm$0.434&6.787$\pm$1.585 &&     ...       &      ...      &      ...      &      ...      \\
log Fe/O (4658)                        &--1.36$\pm$0.03&--1.32$\pm$0.10 &&     ...       &      ...      &      ...      &      ...       \\
                                                                                                                                     \\ 
Fe$^{2+}$/H$^+$, ($\times$10$^7$)(4988)&0.589$\pm$0.026&0.961$\pm$0.131 &&     ...       &      ...      &      ...      &      ...       \\
$ICF$(Fe)                              &      15.08    &      10.53     &&     ...       &      ...      &      ...      &      ...       \\
Fe/H, ($\times$10$^7$)(4988)           &8.875$\pm$0.386&10.120$\pm$1.380&&     ...       &      ...      &      ...      &      ...      \\
log Fe/O (4988)                        &--1.28$\pm$0.02&--1.14$\pm$0.06 &&     ...       &      ...      &      ...      &      ...       \\ 
\hline
\end{tabular}

$^a$ Derived from the fluxes of the [O {\sc iii}] $\lambda$4959, 5007 emission 
lines.

$^b$ Derived from the flux of the  O {\sc ii} $\lambda$4650 recombination
emission line.

\end{table*}

\begin{table*}[t]
\caption{Ionic and total element abundances (UVES observations) \label{tab7}}
\begin{tabular}{lcccccccc} \hline
 &  \multicolumn{6}{c}{0335--052E} & & \\ \cline{2-7}  
Property                        & No. 1+2$^a$     &  No. 1+2$^b$     &   No. 4+5$^b$   &   No. 4+5$^c$&No. (1+2)+(4+5)$^b$& No. 7$^c$      && 0335--052W$^b$\\ \hline\hline

$T_e$(O {\sc iii}), K           &20487$\pm$240  &20585$\pm$246   &19915$\pm$283  &21386$\pm$349  &20158$\pm$239 &18892$\pm$945 &&18196$\pm$390\\
$T_e$(O {\sc ii}), K            &16349$\pm$169  &16360$\pm$172   &16270$\pm$205  &16409$\pm$232  &16308$\pm$171 &16049$\pm$719 &&15839$\pm$306\\  
$T_e$(S {\sc iii}), K           &18704$\pm$199  &18786$\pm$204   &18229$\pm$235  &19450$\pm$290  &18431$\pm$199 &17381$\pm$784 &&16802$\pm$324  \\
$N_e$([S {\sc ii}]), cm$^{-3}$  &     100       &      100       &     100       &218$\pm$65    &     100      &      100      &&  100 \\ 
                                                                                                                                     \\ 
O$^+$/H$^+$, ($\times$10$^5$)     &0.174$\pm$0.005&0.190$\pm$0.006 &0.171$\pm$0.006&0.179$\pm$0.007&0.185$\pm$0.006&0.139$\pm$0.015&&0.400$\pm$0.020 \\
O$^{2+}$/H$^+$, ($\times$10$^5$)  &1.684$\pm$0.048&1.668$\pm$0.048 &1.936$\pm$0.066&1.661$\pm$0.062&1.832$\pm$0.053&1.593$\pm$0.185&&0.959$\pm$0.050 \\
O$^{3+}$/H$^+$, ($\times$10$^6$)  &0.562$\pm$0.021&     ...        &      ...      &0.908$\pm$0.050&      ...      &      ...      &&      ...       \\
O/H, ($\times$10$^5$)             &1.914$\pm$0.048&1.859$\pm$0.049 &2.106$\pm$0.066&1.931$\pm$0.063&2.017$\pm$0.053&1.732$\pm$0.186&&1.359$\pm$0.054 \\
12+log O/H                        &7.28$\pm$0.01  &7.27$\pm$0.01   &7.32$\pm$0.01  &7.28$\pm$0.01  &7.30$\pm$0.01  &7.24$\pm$0.05  &&7.13$\pm$0.02 \\ 
                                                                                                                                                     \\ 
Ne$^{2+}$/H$^+$, ($\times$10$^6$) &2.999$\pm$0.085&2.776$\pm$0.080 &2.782$\pm$0.096&2.746$\pm$0.102&2.906$\pm$0.084&1.774$\pm$0.212 &&0.297$\pm$0.127 \\
$ICF$(Ne)                         &      1.04     &      1.03      &     1.03      &     1.04      &     1.03      &     1.03       &&     1.03       \\
Ne/H, ($\times$10$^6$)            &3.117$\pm$0.097&2.871$\pm$0.089 &2.862$\pm$0.104&2.870$\pm$0.119&2.998$\pm$0.093&1.824$\pm$0.231 &&0.297$\pm$0.127 \\
log Ne/O                          &--0.79$\pm$0.02&--0.81$\pm$0.02 &--0.87$\pm$0.02&--0.83$\pm$0.02&--0.83$\pm$0.02&--0.98$\pm$0.07 &&--0.98$\pm$0.07  \\
                                                                                                                                                     \\ 
N$^{+}$/H$^+$, ($\times$10$^7$)   &0.590$\pm$0.022&0.740$\pm$0.017 &0.500$\pm$0.018&0.532$\pm$0.027&0.680$\pm$0.017&0.496$\pm$0.080 &&0.297$\pm$0.127 \\
$ICF$(N)                          &      10.28    &      9.18      &    11.47      &    10.09      &    10.20      &    11.58       &&    11.58       \\
N/H, ($\times$10$^7$)             &6.059$\pm$0.238&6.795$\pm$0.016 &5.735$\pm$0.217&5.373$\pm$0.288&6.933$\pm$0.185&5.751$\pm$1.001 &&0.297$\pm$0.127 \\
log N/O                           &--1.50$\pm$0.02&--1.44$\pm$0.02 &--1.56$\pm$0.02&--1.56$\pm$0.03&--1.46$\pm$0.02&--1.48$\pm$0.09 &&--1.48$\pm$0.09 \\
                                                                                                                                                     \\ 
S$^{+}$/H$^+$, ($\times$10$^7$)   &0.267$\pm$0.007&0.166$\pm$0.004 &0.152$\pm$0.004&0.248$\pm$0.007&0.165$\pm$0.004&      ...      &&0.435$\pm$0.014 \\
S$^{2+}$/H$^+$, ($\times$10$^7$)  &1.521$\pm$0.100&1.829$\pm$0.061 &1.657$\pm$0.085&1.601$\pm$0.155&1.857$\pm$0.075&      ...      &&2.713$\pm$0.215\\
$ICF$(S)                          &      2.22     &      2.03      &      2.41     &      2.19     &      2.20     &      ...      &&      1.10      \\
S/H, ($\times$10$^7$)             &3.963$\pm$0.223&4.060$\pm$0.124 &4.368$\pm$0.206&4.043$\pm$0.338&4.457$\pm$0.166&      ...      &&3.474$\pm$0.238 \\
log S/O                           &--1.68$\pm$0.03&--1.66$\pm$0.02 &--1.68$\pm$0.02&--1.68$\pm$0.04&--1.66$\pm$0.02&      ...      &&--1.59$\pm$0.03 \\
                                                                                                                                                     \\ 
Cl$^{2+}$/H$^+$, ($\times$10$^8$) &0.298$\pm$0.056&0.272$\pm$0.013&     ...       &      ...     &0.317$\pm$0.024  &      ...      &&      ...    \\
$ICF$(Cl)                         &      1.38     &      1.32     &     ...       &      ...     &      1.38       &      ...      &&      ...     \\
Cl/H, ($\times$10$^8$)            &0.410$\pm$0.078&0.361$\pm$0.017&     ...       &      ...     &0.436$\pm$0.033  &      ...      &&      ...    \\
log Cl/O                          &--3.67$\pm$0.08&--3.71$\pm$0.02&     ...       &      ...     &--3.66$\pm$0.04  &      ...      &&      ...    \\ \\
Ar$^{2+}$/H$^+$, ($\times$10$^7$) &0.525$\pm$0.011&     ...        &     ...       &0.383$\pm$0.012&      ...     &0.385$\pm$0.040 &&      ...    \\
Ar$^{3+}$/H$^+$, ($\times$10$^7$) &0.451$\pm$0.015&     ...        &     ...       &0.495$\pm$0.026&      ...     &       ...      &&      ...     \\
$ICF$(Ar)                         &      2.13     &     ...        &     ...       &      2.10     &      ...     &      2.35      &&      ...     \\
Ar/H, ($\times$10$^7$)            &1.120$\pm$0.039&     ...        &     ...       &0.806$\pm$0.062&      ...     &0.904$\pm$0.095 &&      ...     \\
log Ar/O                          &--2.23$\pm$0.02&     ...        &     ...       &--2.38$\pm$0.04&      ...     &--2.28$\pm$0.06 &&      ...     \\
                                                                                                                                                     \\ 
Fe$^{2+}$/H$^+$, ($\times$10$^7$)(4658)&0.526$\pm$0.026&     ...       &    ...        &      ...     &      ...      &      ...   &&      ...    \\
$ICF$(Fe)                              &      15.18    &     ...       &     ...       &      ...     &      ...      &      ...   &&      ...    \\
Fe/H, ($\times$10$^7$)(4658)           &7.986$\pm$0.397&     ...       &     ...       &      ...     &      ...      &      ...   &&      ...    \\
log Fe/O (4658)                        &--1.38$\pm$0.02&     ...       &     ...       &      ...     &      ...      &      ...   &&      ...    \\
                                                                                                                                                      \\ 
Fe$^{2+}$/H$^+$, ($\times$10$^7$)(4988)&      ...      &     ...       &     ...       &      ...     &      ...      &      ...   &&      ...    \\
$ICF$(Fe)                              &      ...      &     ...       &     ...       &      ...     &      ...      &      ...   &&      ...    \\
Fe/H, ($\times$10$^7$)(4988)           &      ...      &     ...       &     ...       &      ...     &      ...      &      ...   &&      ...    \\
log Fe/O (4988)                        &      ...      &     ...       &     ...       &      ...     &      ...      &      ...   &&      ...    \\
\hline
\end{tabular}

$^a$ ESO program 71.B-0055(A). \\
$^b$ ESO program 70.B-0717(A).  \\
$^c$ ESO program 68.B-0310(A). 

\end{table*}

\begin{table*}[t]
\caption{Helium abundances \label{tab8}}
\begin{tabular}{lccccccc} \hline
 &  \multicolumn{6}{c}{0335--052E}  & \\ \cline{2-7}  
Property                           &No.1+2$^a$&No.1+2$^b$&No.1+2$^c$&No.4+5$^c$&No.7$^a$&No.7$^b$& 0335--052W$^b$\\ \hline\hline
\multicolumn{7}{c}{a) Benjamin et al. (2002) emissivities}  \\ \hline
$\Delta$$I$(H$\alpha$)/$I$(H$\alpha$)& 0.0004          & 0.0013          & 0.0177          & 0.0007          & 0.0500          & 0.0491          & 0.0490           \\
$T_e$(He$^+$)                        & 19520           & 20250           & 20410           & 20330           & 19110           & 18970           & 17030            \\
$N_e$(He$^+$)                        &   162           &   132           &    53           &    12           &    12           &    10           &    14            \\
$\tau$($\lambda$3889)                &   4.59          &   5.01          &   5.01          &   1.24          &   0.01          &   0.02          &   0.02           \\
$y^+$($\lambda$4471)                 &0.0834$\pm$0.0021&0.0820$\pm$0.0013&0.0827$\pm$0.0013&0.0834$\pm$0.0028&0.0858$\pm$0.0118&0.0812$\pm$0.0030&0.0935$\pm$0.0086 \\
$y^+$($\lambda$5876)                 &0.0852$\pm$0.0014&0.0839$\pm$0.0013&0.0777$\pm$0.0012&0.0798$\pm$0.0015&0.0816$\pm$0.0035&0.0802$\pm$0.0020&0.0780$\pm$0.0065 \\
$y^+$($\lambda$6678)                 &0.0768$\pm$0.0017&0.0798$\pm$0.0014&0.0818$\pm$0.0017&0.0792$\pm$0.0016&0.0876$\pm$0.0092&0.0812$\pm$0.0049&0.0791$\pm$0.0137 \\
$y^+_{wm}$                          &0.0822$\pm$0.0010&0.0821$\pm$0.0008&0.0805$\pm$0.0008&0.0801$\pm$0.0010&0.0826$\pm$0.0031&0.0806$\pm$0.0016&0.0831$\pm$0.0048 \\
$y^{2+}$($\lambda$4686)              &0.0011$\pm$0.0001&0.0012$\pm$0.0000&0.0026$\pm$0.0000&0.0041$\pm$0.0002&        ...      &0.0012$\pm$0.0001&        ...       \\
$ICF$                                &0.9936           &0.9935           &0.9933           &0.9928           &0.9928           &0.9929           &0.9896            \\
$y$($\lambda$4471)                   &0.0839$\pm$0.0021&0.0827$\pm$0.0013&0.0848$\pm$0.0013&0.0868$\pm$0.0028&0.0852$\pm$0.0117&0.0818$\pm$0.0030&0.0926$\pm$0.0085 \\
$y$($\lambda$5876)                   &0.0857$\pm$0.0014&0.0846$\pm$0.0013&0.0798$\pm$0.0012&0.0833$\pm$0.0015&0.0811$\pm$0.0035&0.0809$\pm$0.0020&0.0772$\pm$0.0064 \\
$y$($\lambda$6678)                   &0.0774$\pm$0.0017&0.0805$\pm$0.0014&0.0838$\pm$0.0017&0.0827$\pm$0.0016&0.0870$\pm$0.0092&0.0819$\pm$0.0049&0.0783$\pm$0.0135 \\
$y_{wm}$                            &0.0828$\pm$0.0010&0.0828$\pm$0.0008&0.0825$\pm$0.0008&0.0835$\pm$0.0010&0.0820$\pm$0.0031&0.0812$\pm$0.0016&0.0822$\pm$0.0048 \\
$Y$                            &0.2486$\pm$0.0030&0.2486$\pm$0.0024&0.2481$\pm$0.0024&0.2503$\pm$0.0032&0.2470$\pm$0.0097&0.2452$\pm$0.0049&0.2474$\pm$0.0148 \\ \hline
\multicolumn{7}{c}{b) Porter et al. (2005) emissivities}    \\ \hline
$\Delta$$I$(H$\alpha$)/$I$(H$\alpha$)& 0.0010          & 0.0027          & 0.0322          & 0.0002          & 0.0488          & 0.0493          & 0.0497           \\
$T_e$(He$^+$)                        & 19520           & 20260           & 20360           & 20330           & 19060           & 18850           & 16910            \\
$N_e$(He$^+$)                        &   197           &   162           &    86           &    10           &    11           &    11           &    11            \\
$\tau$($\lambda$3889)                &   4.52          &   5.00          &   5.01          &   1.44          &   0.02          &   0.02          &   0.04           \\
$y^+$($\lambda$4471)                 &0.0853$\pm$0.0021&0.0841$\pm$0.0013&0.0849$\pm$0.0013&0.0862$\pm$0.0028&0.0884$\pm$0.0118&0.0835$\pm$0.0030&0.0958$\pm$0.0086 \\
$y^+$($\lambda$5876)                 &0.0860$\pm$0.0014&0.0848$\pm$0.0013&0.0792$\pm$0.0012&0.0816$\pm$0.0015&0.0835$\pm$0.0035&0.0820$\pm$0.0020&0.0800$\pm$0.0065 \\
$y^+$($\lambda$6678)                 &0.0771$\pm$0.0017&0.0800$\pm$0.0014&0.0831$\pm$0.0017&0.0796$\pm$0.0016&0.0882$\pm$0.0092&0.0818$\pm$0.0049&0.0803$\pm$0.0137 \\
$y^+_{wm}$                          &0.0831$\pm$0.0010&0.0831$\pm$0.0008&0.0821$\pm$0.0008&0.0814$\pm$0.0010&0.0844$\pm$0.0031&0.0824$\pm$0.0016&0.0851$\pm$0.0048 \\
$y^{2+}$($\lambda$4686)              &0.0011$\pm$0.0001&0.0012$\pm$0.0000&0.0026$\pm$0.0000&0.0041$\pm$0.0002&       ...       &0.0012$\pm$0.0001&       ...        \\
$ICF$                                &0.9936           &0.9935           &0.9933           &0.9928           &0.9928           &0.9929           &0.9896            \\
$y$($\lambda$4471)                   &0.0858$\pm$0.0021&0.0848$\pm$0.0013&0.0869$\pm$0.0013&0.0897$\pm$0.0028&0.0877$\pm$0.0117&0.0842$\pm$0.0030&0.0948$\pm$0.0085 \\
$y$($\lambda$5876)                   &0.0865$\pm$0.0013&0.0854$\pm$0.0012&0.0812$\pm$0.0012&0.0851$\pm$0.0015&0.0829$\pm$0.0035&0.0827$\pm$0.0020&0.0792$\pm$0.0064 \\
$y$($\lambda$6678)                   &0.0776$\pm$0.0017&0.0807$\pm$0.0014&0.0851$\pm$0.0017&0.0831$\pm$0.0016&0.0876$\pm$0.0091&0.0825$\pm$0.0049&0.0794$\pm$0.0135 \\
$y_{wm}$                            &0.0836$\pm$0.0009&0.0838$\pm$0.0008&0.0842$\pm$0.0008&0.0849$\pm$0.0010&0.0838$\pm$0.0031&0.0831$\pm$0.0016&0.0842$\pm$0.0048 \\
$Y$                            &0.2506$\pm$0.0029&0.2510$\pm$0.0023&0.2518$\pm$0.0024&0.2533$\pm$0.0031&0.2509$\pm$0.0096&0.2493$\pm$0.0048&0.2518$\pm$0.0148 \\ \hline
\end{tabular}

$^a$ low-resolution FORS observations. \\
$^b$ high-resolution FORS observations. \\
$^c$ UVES observations. \\
\end{table*}

\end{document}